\documentclass[journal]{IEEEtran}
\usepackage{amssymb,graphicx,subfigure,latexsym,fullpage}
\newlength{\bottomwidth}
\newlength{\mypicheight}
\newlength{\mytextheight}
\newsavebox{\mypicbox}

\newcommand{\picbox}[2]{\savebox{\mypicbox}{\includegraphics[width=#1]{#2}}\settoheight{\mypicheight}{\usebox{\mypicbox}}\settoheight{\mytextheight}{W}\addtolength{\mypicheight}{-\mytextheight}\raisebox{-\mypicheight}{\usebox{\mypicbox}}}

\newcommand{\sourcewords}{\ensuremath{\mathcal{X}}}
\newcommand{\representations}{\ensuremath{\mathcal{Y}}}

\newcommand{\reals}{\ensuremath{\mathbb{R}}}

\newcommand{\pool}{\ensuremath{\mathcal{P}}}

\newcommand{\tuple}[2]{\ensuremath{\left\langle#1,#2\right\rangle}}
\newcommand{\lte}{\ensuremath{\preccurlyeq}}

\newcommand{\mrk}[2]{\settowidth{\bottomwidth}{#2}\makebox(0,0)[b]{\smash{\kern\bottomwidth\raise.8em#1}}#2}

\newcommand{\markdeletion}{\tiny\ensuremath{\boxdot}}

\newcommand{\markplus}[1]{\mrk{\hbox{\tiny\ensuremath{\blacktriangle}}}{\hbox{#1}}}
\newcommand{\markminus}[1]{\mrk{\hbox{\tiny\ensuremath{\blacktriangledown}}}{\hbox{#1}}}
\newcommand{\marksame}[1]{\mrk{\hbox{\scriptsize\ensuremath{\bullet}}}{\hbox{#1}}}
\newcommand{\markchange}[1]{\mrk{\hbox{\tiny\ensuremath{\bigstar}}}{\hbox{#1}}}

\newcommand{\smi}[1]{\includegraphics[width=0.41\textwidth, trim=1 2 2
    1]{#1}}
\newcommand{\smj}[1]{\quad\includegraphics[width=0.32\textwidth, trim=1 2 2
    1]{#1}\quad}

\newcommand{\floor}[1]{\left\lfloor{#1}\right\rfloor}

\newcommand{\object}[1]{\ensuremath{\mathbb#1}}
\newcommand{\os}[2]{\ensuremath{#1_\textsc{\footnotesize #2}}}
\newcommand{\obj}{\object{O}}
\newcommand{\mouse}{\object{M}}
\newcommand{\cross}{\object{C}}
\newcommand{\nmouse}{\object{N}}
\newcommand{\wilde}{\object{W}}

\begin{document}

\title{Approximating Rate-Distortion Graphs of Individual Data: Experiments
in Lossy Compression and Denoising}
\author{Steven de Rooij and Paul Vit\'anyi\thanks{CWI, INS4, Kruislaan
    413, P.O. Box 94079, 1090 GB Amsterdam.}\thanks{Corresponding
    author: Steven de Rooij, \texttt{rooij@cwi.nl}}}
\date{\today}
\maketitle

\begin{abstract}
Classical rate-distortion theory requires knowledge of an elusive
source distribution.  Instead, we analyze rate-distortion properties
of individual objects using the recently developed algorithmic
rate-distortion theory.  The latter is based on the noncomputable
notion of Kolmogorov complexity.  To apply the theory we approximate
the Kolmogorov complexity by standard data compression techniques, and
perform a number of experiments with lossy compression and denoising
of objects from different domains.  We also introduce a natural
generalization to lossy compression with side information. To maintain
full generality we need to address a difficult searching
problem. While our solutions are therefore not time efficient, we do
observe good denoising and compression performance.
\end{abstract}

\begin{keywords}
compression, denoising, rate-distortion, structure function,
Kolmogorov complexity
\end{keywords}

\section{Introduction}
Rate-distortion theory analyzes communication over a channel under a
constraint on the number of transmitted bits, the ``rate''. It
currently serves as the theoretical underpinning for many important
applications such as lossy compression and denoising, or more
generally, applications that require a separation of structure and
noise in the input data.

Classical rate-distortion theory evolved from Shannon's theory of
communication\cite{Shannon1948}. It studies the trade-off between the
rate and the achievable fidelity of the transmitted representation
under some distortion function, where the analysis is carried out
\emph{in expectation} under some source distribution. Therefore the
theory can only be meaningfully applied if we have some reasonable
idea as to the distribution on objects that we want to compress
lossily. While lossy compression is ubiquitous, propositions with
regard to the underlying distribution tend to be ad-hoc, and
necessarily so, because (1) it is a questionable assumption that the
objects that we submit to lossy compression are all drawn from the
same probability distribution, or indeed that they are drawn from a
distribution at all, and (2) even if a true source distribution is
known to exist, in most applications the sample space is so large that
it is extremely hard to determine what it is like: objects that occur
in practice very often exhibit more structure than predicted by the
used source model.

For large outcome spaces then, it becomes important to consider
structural properties of \emph{individual objects}.  For example, if
the rate is low, then we may still be able to transmit objects that
have a very regular structure without introducing any distortion, but
this becomes impossible for objects with high information density.
This point of view underlies some recent research in the lossy
compression community\cite{Scheirer2001}. At about
the same time, a rate-distortion theory which allows analysis of
individual objects has been developed within the framework of
Kolmogorov complexity\cite{LiVitanyi1997}. It defines a
rate-distortion function not with respect to some elusive source
distribution, but with respect to an individual source word. Every
source word thus obtains its own associated rate-distortion function.

We will first give a brief intoduction to algorithmic rate-distortion
theory in Section~\ref{sec:algorithmic}. We also describe a novel
generalization of the theory to settings with side information, and we
describe two distinct applications of the theory, namely lossy
compression and denoising.

Algorithmic rate-distortion theory is based on Kolmogorov complexity,
which is not computable. We nevertheless cross the bridge between
theory and practice in Section~\ref{sec:practical}, by approximating
Kolmogorov complexity by the compressed size of the object by
 a general purpose data
compression algorithm. Even so, approximating the rate-distortion
function is a difficult search problem. We motivate and outline the
genetic algorithm we used to approximate the rate-distortion function.

In Section~\ref{sec:experiments} we describe four experiments in lossy
compression and denoising. The results are presented and discussed in
Section~\ref{sec:resdisc}. Then, in Section~\ref{sec:quality} we take
a step back and discuss to what extent our practical approach yields a
faithful approximation of the theoretical algorithmic rate-distortion
function. We end with a conclusion in Section~\ref{sec:conclusion}.

\section{Algorithmic Rate-Distortion}\label{sec:algorithmic}
Suppose we want to communicate objects $x$ from a set of source words
$\sourcewords$ using at most $r$ bits per object. We call $r$ the
\emph{rate}. We locate a good \emph{representation} of $x$ within a
finite set $\representations$, which may be different from
$\sourcewords$ in general (but we usually have
$\sourcewords=\representations$ in this text). The lack of fidelity of
a representation $y$ is quantified by a distortion function
$d:\sourcewords\times\representations\rightarrow\reals$.

\begin{figure}[!ht]
\centerline{\includegraphics[width=\columnwidth]{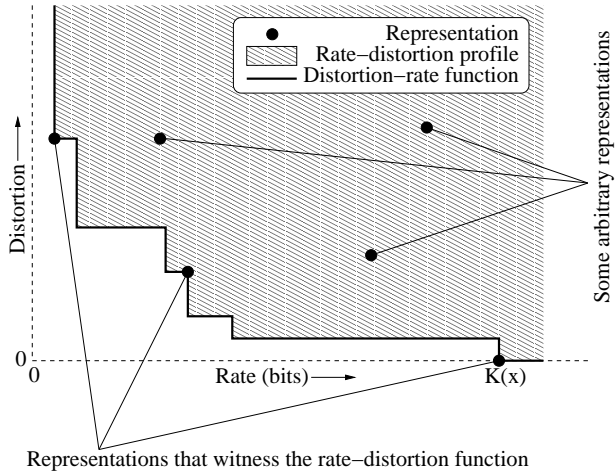}}
\caption{Rate-distortion profile and distortion-rate function}
\label{fig:profile}
\end{figure}

The Kolmogorov complexity of $y$, denoted $K(y)$, is the length of the
shortest program that constructs $y$. More precisely, it is the length
of the shortest input to a fixed universal binary prefix machine that
will output $y$ and then halt; also see the
textbook\cite{LiVitanyi1997}. We can transmit any representation $y$
that has $K(y)\le r$, the receiver can then run the program to obtain
$y$ and is thus able to reconstruct $x$ up to distortion $d(x,y)$.
Define the \emph{rate-distortion profile $P_x$ of the source word
$x$} as the set of pairs $\tuple{r}{a}$ such that there is a
representation $y\in\representations$ with $d(x,y)\le a$ and $K(y)\le
r$. The possible combinations of $r$ and $a$ can also be characterised
by the \emph{rate-distortion function of the source word $x$}, which
is defined as $r_x(a)=\min\{r:\tuple{r}{a}\in P_x\}$, or by the
\emph{distortion-rate function of the source word $x$}, which is
defined as $d_x(r)=\min\{a:\tuple{r}{a}\in P_x\}$. These two functions
are somewhat like inverses of each other; although strictly speaking
they are not since they are monotonic but not strictly monotonic. A
representation $y$ is said to \emph{witness} the rate-distortion
function of $x$ if $r_x(d(x,y))=K(y)$. These definitions are
illustrated in Figure~\ref{fig:profile}.

Algorithmic rate-distortion theory is developed and treated in much
more detail in~\cite{VereshchaginVitanyi2005}. It is a generalization
of Kolmogorov's structure function theory,
see~\cite{VereshchaginVitanyi2004}.

\subsection{Side Information}\label{sec:sideinfo}
We generalize the algorithmic rate-distortion framework, so that it
can accommodate side information. Suppose that we want to transmit a
source word $x\in\sourcewords$ and we have chosen a representation
$y\in\representations$ as before. The encoder and decoder often share
a lot of information: both might know that grass is green and the sky
is blue, they might share a common language, and so on. They would not
need to transmit such information. If encoder and decoder share some
information $z$, then the programs they transmit to compute the
representation $y$ may use this side information $z$. Such programs
might be shorter than they could be otherwise.  This can be formalised
by switching to the \emph{conditional} Kolmogorov complexity $K(y|z)$,
which is the length of the shortest Turing machine program that
constructs $y$ on input $z$. We redefine $K(y)=K(y|\epsilon)$, where
$\epsilon$ is the empty sequence, so that $K(y|z)\le K(y)+O(1)$: the
length of the shortest program for $y$ can never significantly
increase when side information is provided, but it might certainly
decrease when $y$ and $z$ share a lot of information\cite{LiVitanyi1997}. We change the definitions as follows: The
\emph{rate-distortion profile of the source word $x$ with side
information $z$} is the set of pairs $\tuple{r}{a}$ such that there is
a representation $y\in\representations$ with $d(x,y)\le a$ and
$K(y|z)\le r$. The definitions of the rate-distortion function and the
distortion-rate function are similarly changed. Henceforth we will
omit mention of the side information $z$ unless it is relevant to the
discussion.

While this generalization seems very natural the authors are not aware
of earlier proposals along these lines. In
Section~\ref{sec:resdisc} we will demonstrate one use for this
generalized rate-distortion theory: removal of spelling errors in
written text, an example where denoising is not practical without use
of side information.

\subsection{Distortion Spheres and the Minimal Sufficient Statistic}
\label{sec:sufstat}
A representation $y$ that witnesses the rate-distortion function is
the best possible rendering of the source object $x$ at the given rate
because it minimizes the distortion, but if the rate is lower than
$K(x)$, then some information is necessarily lost. Since one of our
goals is to find the best possible separation between structure and
noise in the data, it is important to determine to what extent the
discarded information is noise.

Given a representation $y$ and the distortion $a=d(x,y)$, we can find
the source object $x$ somewhere on the list of all $x'\in\sourcewords$
that satisfy $d(x',y)=a$. The information conveyed about $x$ by $y$
and $a$ is precisely, that $x$ can be found on this list. We call such
a list a \emph{distortion sphere}. A distortion sphere of radius $a$,
centred around $y$ is defined as follows:
\begin{equation}
S_y(a):=\{x'\in\sourcewords:d(x',y)=a\}.\label{eq:sphere}
\end{equation}
If the discarded information is pure white noise, then this means that
$x$ must be a completely random element of this list.
Conversely, all random elements in the list share all ``simply described''
(in the sense of having low Kolmogorov complexity) properties
that $x$ satisfies. Hence, with respect to the ``simply described''
properties, every such random element is as good as $x$,
see \cite{VereshchaginVitanyi2005} for more details. In such cases a
literal specification of the index of $x$ (or any other random
element) in the list is the most
efficient code for $x$, given only that it is in $S_y(a)$. A
fixed-length, literal code requires $\log|S_y(a)|$ bits. (Here and in
the following, all logarithms are taken to base $2$ unless otherwise
indicated.) On the other hand, if the discarded information is
structured, then the Kolmogorov complexity of the index of $x$ in
$S_y(a)$ will be significantly lower than the logarithm of the size of
the sphere.  The difference between these two codelengths can be used
as an indicator of the amount of structural information that is
discarded by the representation $y$.  Vereshchagin and
Vit\'anyi\cite{VereshchaginVitanyi2005} call this quantity the
\emph{randomness deficiency} of the source object $x$ in the set
$S_y(a)$, and they show that if $y$ witnesses the rate-distortion
function of $x$, then it \emph{minimizes} the randomness deficiency at
rate $K(y)$; thus the rate-distortion function identifies those
representations that account for as much structure as possible at the
given rate.

To assess how much structure is being discarded at a given rate,
consider a code for the source object $x$ in which we first transmit
the shortest possible program that constructs both a representation
$y$ and the distortion $d(x,y)$, followed by a literal, fixed-length
index of $x$ in the distortion sphere $S_y(a)$. Such a code has the
following length function:
\begin{equation}
L_y(x):=K(\tuple{y}{d(x,y)})+\log|S_y(d(x,y))|.
\label{eq:codelength}
\end{equation}
If the rate is very low then the representation $y$ models only very
basic structure and the randomness deficiency in the distortion sphere
around $y$ is high. Borrowing terminology from statistics, we may say
that $y$ is a representation that ``underfits'' the data. In such
cases we should find that $L_y(x)>K(x)$, because the fixed-length code
for the index of $x$ within the distortion sphere is suboptimal in
this case. But suppose that $y$ is complex enough that it satisfies
$L_y(x)\approx K(x)$. In \cite{VereshchaginVitanyi2005}, such
representations are called \emph{(algorithmic) sufficient statistics}
for the data $x$. A sufficient statistic has close to zero randomness
deficiency, which means that it represents all structure that can be
detected in the data. However, sufficient statistics might contain not
only structure, but noise as well. Such a representation would be
overly complex, an example of overfitting. A \emph{minimal} sufficient
statistic balances between underfitting and overfitting. It is defined
as the lowest complexity sufficient statistic, which is the same as
the lowest complexity representation $y$ that minimizes the codelength
$L_y(x)$. As such it can also be regarded as the ``model'' that should
be selected on the basis of the Minimum Description Length (MDL)
principle\cite{BarronRY98}. To be able to relate the distortion-rate
function to this codelength we define the \emph{codelength function}
$\lambda_x(r)=L_y(x)$ where $y$ is the representation that minimizes
the distortion at rate $r$.\footnote{This is superficially similar to
the MDL function defined in~\cite{VereshchaginVitanyi2004}, but it is
\emph{not} exactly the same since it involves optimisation of the
distortion at a given rate rather than direct optimisation of the code
length.}

\subsection{Applications: Denoising and Lossy Compression}\label{sec:apps}
Representations that witness the rate-distortion function
provide optimal separation between structure that can be expressed
at the given rate and residual information that is perceived as noise
Therefore, these representations can be interpreted as denoised versions of the
original. In denoising, the goal is of course to discard as much noise
as possible, without losing any structure. Therefore the minimal
sufficient statistic, which was described in the previous section, is
the best candidate for applications of denoising.

While the minimum sufficient statistic is a denoised representation of
the original signal, it is not necessarily given in a directly usable
form. For instance, $\representations$ could consist of subsets of
$\sourcewords$, but a \emph{set} of source-words is not always
acceptable as a denoising result. So in general one may need to apply
some function $f:\representations\rightarrow\sourcewords$ to the
sufficient statistic to construct a usable object. But if
$\sourcewords=\representations$ and the distortion function is a
metric, as in our case, then the representations are already in an
acceptable format, so here we use the identity function for the
transformation $f$.

In applications of lossy compression, one may be willing to accept a
rate which is lower than the minimal sufficient statistic complexity,
thereby losing some structural information. However, for a minimal
sufficient statistic $y$, theory does tell us that it is not
worthwhile to set the rate to a higher value than the complexity of
$y$. The original object $x$ is a random element of
$S_y(d(x,y))$, and it cannot be distinguished from any
other random $z\in S_y(d(x,y))$ using only ``simply described'' properties.
So we have no ``simply described'' test to discredit the hypothesis
that $x$ (or any such $z$) is the original object, given $y$
and $d(x,y)$. But increasing the rate, yielding a model $y'$ and
$d(x,y') < d(x,y)$, we commonly obtain a sphere $S_{y'}$ of smaller
cardinality than $S_y$, with some random elements of $S_y$ not
being random elements of $S_{y'}$. These excluded elements, however,
were perfectly good candidates of being the original object. That is,
at rate higher than that of the minimal sufficient statistic,
the resulting representation $y'$ models irrelevant features that
are specific to $x$, that is, noise and no structure, that exclude
viable candidates for the olriginal object: the representation
starts to ``overfit''.

In lossy compression, as in denoising, the representations themselves
may be unsuitable for presentation to the user. For example, when
decompressing a lossily compressed image, in most applications a
\emph{set} of images would not be an acceptable result. So again a
transformation from representations to objects of a usable form has to
be specified. There are two obvious ways of doing this:
\begin{enumerate}
\item If a representation $y$ witnesses the rate-distortion function
  for a source word $x\in\sourcewords$, then this means that $x$
  cannot be distinguished from any other object $x'\in S_y(d(x,y))$ at
  rate $K(y)$. Therefore we should not use a deterministic
  transformation, but rather report the uniform distribution on
  $S_y(d(x,y))$ as the lossily compressed version of $x$. This method
  has the advantage that it is applicable whether or not
  $\sourcewords=\representations$.
\item On the other hand, if $\sourcewords=\representations$ and the
  distortion function is a metric, then it makes sense to use the
  identity transformation again, although here the motivation is
  different. Suppose we select some $x'\in S_y(d(x,y))$ instead of
  $y$. Then the best upper bound we can give on the distortion is
  $d(x,x')\le d(x,y)+d(y,x')=2d(x,y)$ (by the triangle inequality and
  symmetry). On the other hand if we select $y$, then the distortion
  is exactly $d(x,y)$, which is only half of the upper bound we
  obtained for $x'$. Therefore it is more suitable if one adopts a
  worst-case approach. This method has as an additional advantage that
  the decoder does not need to \emph{know} the distortion $d(x,y)$
  which often cannot be computed from $y$ without knowledge of $x$.
\end{enumerate}
To illustrate the difference one may expect from these approaches,
consider the situation where the rate is lower than the rate that
would be required to specify a sufficient statistic. Then intuitively,
all the noise in the source word $x$ as well as some of the structure
are lost by compressing it to a representation $y$. The second method
immediately reports $y$, which contains a lot less noise than the
source object $x$; thus $x$ and $y$ are qualitatively different, which
may be undesirable. On the other hand, the compression result will be
qualitatively different from $x$ anyway, because the rate simply is
too low to retain all structure. If one would apply the first
approach, then a result $x'$ would likely contain \emph{more} noise
than the original, because it contains less structure at the same
level of distortion (meaning that $K(x')>K(x)$ while
$d(x',y)=d(x,y)$).

If the rate is high enough to transmit a sufficient statistic, then
the first approach seems preferable. We have nevertheless chosen to
always report $y$ directly in our analysis, which has the advantage
that this way, all reported results are of the same type.

\section{Computing Individual Object Rate-Distortion}\label{sec:practical}
The rate-distortion function for an object $x$ with side information
$z$ and a distortion function $d$ is found by simultaneous
minimizing two objective functions
\begin{eqnarray}
\label{eq.gfunctions}
g_1(y) & = & K(y|z)
\\ g_2(y) & = & d(x,y)
\\ g(y) &=& \tuple{g_1(y)}{g_2(y)}. 
\nonumber
\end{eqnarray}
We call the tuple $g(y)$ the
\emph{trade-off} of $y$. We impose a partial order on representations:
\begin{equation}\label{eq.leq}
y\lte y'
\end{equation}
 if and only if $g_1(y)\le g_1(y')$ and $g_2(y)\le
g_2(y')$. Our goal is to find the set of Pareto-optimal
representations, that is, the set of representations that are minimal
under $\lte$.

Such an optimisation problem cannot be implemented because of the
uncomputability of $K(\cdot)$. To make the idea practical, we need to
approximate the conditional Kolmogorov complexity. As observed in
\cite{VitanyiCilibrasi2004}, it follows directly from symmetry of
information for Kolmogorov complexity (see
\cite[p.233]{LiVitanyi1997}) that:
\begin{equation}
K(y|z)=K(zy)-K(z)+O(\log n),
\label{eq:kcond}
\end{equation}
where $n$ is the length of $zy$. Ignoring the logarithmic term, this
quantity can be approximated by replacing $K(\cdot)$ by the length of
the compressed representation under a general purpose compression
algorithm $A:\sourcewords\rightarrow\{0,1\}^*$. The length of the
compressed representation of $x$ is denoted by $L_A(x)$. This way we
obtain, up to an additive independent constant:
\begin{equation}
K(y|z)\leq L_A(zy)-L_A(z).
\label{eq:kapprox}
\end{equation}
We redefine $g_1(y):=L_A(zy)-L_A(z)$ in order to get a
practical objective function.

This may be a poor approximation: we only have that $0\le K(y|z)\le
L_A(zy)-L_A(z)$, up to a constant,
 so the compressed size is an upper bound that may be
quite high even for objects that have Kolmogorov complexity close to
zero. Our results show evidence that some of the theoretical
properties of the distortion-rate function nevertheless carry over to
the practical setting; we also explain how some observations that are
not predicted by theory are in fact related to the (unavoidable)
inefficiencies of the used compressor.

\subsection{Compressor (rate function)}\label{sec:compressor}
We could have used any general-purpose compressor in
(\ref{eq:kapprox}), but we chose to implement our own for three
reasons:
\begin{itemize}
\item It should be both fast and efficient. We can gain some advantage
  over other available compressors because there is no need to
  actually construct a code. It suffices to compute code
  \emph{lengths}, which is much easier. As a secondary advantage, the
  codelengths we compute are not necessarily multiples of eight bits:
  we allow rational idealised codelengths, which may improve
  precision.
\item It should not have any arbitrary restrictions or
  optimisations. Most general purpose compressors have limited window
  sizes or optimisations to improve compression of common file types;
  such features could make the results harder to interpret.
\end{itemize}
In our experiments we used a block sorting compression algorithm with
a move-to-front scheme as descibed in \cite{burrowswheeler1994}. In
the encoding stage M2 we employ a simple statistical model and omit
the actual encoding as it suffices to accumulate codelengths. The
source code of our implementation (in C) is available from the authors
upon request. The resulting algorithm is very similar to a number of
common general purpose compressors, such the freely available
bzip2\cite{bzip2} and zzip\cite{zzip}, but it is simpler and faster
for small inputs.

Of course, domain specific compressors might yield better compression
for some object types (such as sound wave files), and therefore a
better approximation of the Kolmogorov complexity. However, the
compressor that we implemented is quite efficient for objects of many
of the types that occur in practice; in particular it compressed the
objects of our experiments (text and small images) quite well. We have
tried to improve compression performance by applying standard image
preprocessing algorithms to the images, but this turned out not to
improve compression at all. Figure~\ref{tab:compression} lists the
compressed size of an image of a mouse under various different
compression and filtering regimes. Compared to other compressors, ours
is quite efficient; this is probably because other compressors are
optimised for larger files and because we avoid all overhead inherent
in the encoding process. Most compressors have optimisations for text
files which might explain why our compressor compares less favourably
on the Oscar Wilde fragment.

\begin{figure*}[!ht]
\begin{center}
\begin{tabular}{l|r|r|r|l}
  Compression&\multicolumn{1}{c|}{mouse}&\multicolumn{1}{c|}{cross}&\multicolumn{1}{c|}{Wilde}&description\\
  \hline
  A&7995.11&3178.63&3234.45&Our compressor, described in~\S\ref{sec:compressor}\\
  \texttt{zzip}&8128.00&3344.00&3184.00&An efficient block sorting compressor\\
  \texttt{PPMd}&8232.00&2896.00&2744.00&High end statistical compressor\\
   RLE $\rightarrow$ A&8341.68&3409.22&--&A with run length encoding filter\\
  \texttt{bzip2}&9296.00&3912.00&3488.00&Widespread block sorting compressor\\
  \texttt{gzip}&9944.00&4008.00&3016.00&LZ77 compressor\\
  sub $\rightarrow$ A&10796.29&4024.26&--&A with Sub filter\\
  paeth $\rightarrow$ A&13289.34&5672.70&--&A with Paeth filter\\
  None&20480.00&4096.00&5864.00&Literal description\\
\end{tabular}
\end{center}
\caption{Compressed sizes of three objects that we experiment
  upon. See Figure~\ref{fig:mouse} for the mouse,
  Figure~\ref{fig:cross} for the cross with added noise and
  Figure~\ref{fig:dorian_gray} for the corrupted Oscar Wilde fragment
  (the middle version). In the latter we give the compressed size
  \emph{conditional} on a training text, like in the
  experiments. ``A'' is our own algorithm, described in
  Section~\ref{sec:compressor}. For a description of the filters
  see~\cite{OReillyPNG}.}
\label{tab:compression}
\end{figure*}

\subsection{Codelength Function}
In Section~\ref{sec:sufstat} we introduced the codelength function
$\lambda_x(r)$. Its definition makes use of (\ref{eq:codelength}), for
which we have not yet provided a computable alternative. We
use the following approximation:
\begin{equation}
L_y(x)\approx L'_y(x)=L_A(y)+L_D(d(x,y)|y)+L_U(x|y,d(x,y)),
\end{equation}
where $L_D$ is yet another code which is necessary to specify the
radius of the distortion sphere around $y$ in which $x$ can be
found. It is possible that this distortion is uniquely determined by
$y$, for example if $\representations$ is the set of all finite
subsets of $\sourcewords$ and list decoding distortion is used, as
described in~\cite{VereshchaginVitanyi2004}. If $d(x,y)$ is a function
of $y$ then $L_D(d(x,y)|y)=0$. In other cases, the representations do
not hold sufficient information to determine the distortion. This is
typically the case when $\sourcewords=\representations$ as in the
examples in this text. In that case we actually need to encode
$d(x,y)$ separately. It turns out that the number of bits that are
required to specify the distortion are negligible in proportion to the
total three part codelength. In the remainder of the paper we use for
$L_D$ a universal code on the integers similar to the one described
in~\cite{LiVitanyi1997}; it has codelength
$L_D(d)=\log(d+1)+O(\log\log d)$.

\subsection{Distortion Functions}\label{sec:distortion}
We use three common distortion functions which we describe below. All
distortion functions used in this text are metrics, which have the
property that $\sourcewords=\representations$.

\paragraph{Hamming distortion}
Hamming distortion is perhaps the simplest distortion function that
could be used. Let $x$ and $y$ be two objects of equal length $n$. The
Hamming distortion $d(x,y)$ is equal to the number of symbols in $x$
that do not match those in the corresponding positions in
$y$.

\paragraph{Euclidean distortion}
As before, let $x=x_1\ldots x_n$ and $y=y_1\ldots y_n$ be two objects
of equal length, but the symbols now have a numerical
interpretation. Euclidean distortion is
$d(x,y)=\sqrt{\sum_{i=1}^n(x_i-y_i)^2}$: the distance between $x$ and
$y$ when they are interpreted as vectors in an $n$-dimensional
Euclidean space. Note that this definition of Euclidean distortion
differs from the one in~\cite{VereshchaginVitanyi2005}.

\paragraph{Edit distortion}
The edit distortion of two strings $x$ and $y$, of possibly different
lengths, is the minimum number of symbols that have to be deleted
from, inserted into, or changed in $x$ in order to obtain $y$ (or vice
versa)\cite{Levenshtein1966}. It is also known as \emph{Levenshtein
distortion}. It is a well-known measure that is often used in
applications that require approximate string matching.

\subsection{Searching for the Rate-Distortion Function}\label{sec:search}
The search problem that we propose to address has two properties that
make it very hard. Firstly, the search space is enormous: for an
object of $n$ bits there are $2^n$ candidate representations of the
same size, and objects that are typically subjected to lossy
compression are often millions or billions of bits long. Secondly, we
want to avoid making too many assumptions about the two objective
functions, so that we can later freely change the compression
algorithm and the distortion function. Under such circumstances the
two most obvious search methods are not practical:

\begin{itemize}
\item An exhaustive search is infeasible for search spaces of such
  large size, unless more specific properties of the objective
  functions are used in the design of the algorithm. To investigate
  how far we could take such an approach, we have implemented an
  exhaustive algorithm under the requirement that, given a prefix of a
  representation $y$, we can compute reasonable lower bounds on the
  values of both objective functions $g_1$ and $g_2$. This allows for
  relatively efficient enumeration of all representations of which the
  objective functions do not exceed specific maxima: it is never
  necessary to consider objects which have a prefix for which the
  lower bounds exceed the constraints, which allows for significant
  pruning. In this fashion we were able to find the rate-distortion
  function under Hamming distortion for objects of which the
  compressed size is about 25 bits or less within a few hours on a
  desk-top computer.
\item A greedy search starts with a poor solution and iteratively
  makes modifications that constitute strict improvements. We found
  that this procedure tends to terminate quickly in some local optimum
  that is very bad globally.
\end{itemize}
Since the structure of the search landscape is at present poorly
understood and we do not want to make any unjustifiable assumptions,
we use a genetic search algorithm which performs well enough that
interesting results can be obtained. It is described in the Appendix
\ref{sect.ga}.

\section{Experiments}\label{sec:experiments}
\begin{figure*}[!ht]
\newlength{\exppicwidth}\exppicwidth=0.38\columnwidth
\footnotesize
\noindent\begin{tabular}{cp{0.56\columnwidth}cp{0.56\columnwidth}}
\picbox{\exppicwidth}{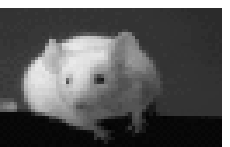}&A picture of a
  mouse of $64\times 40$ pixels. The picture is analyzed with respect
  to Euclidean distortion.%
&\picbox{0.85\exppicwidth}{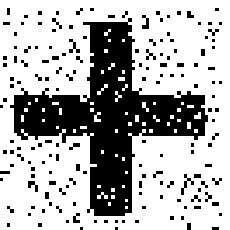}&A noisy
  monochrome image of $64\times 64$ pixels that depicts a cross. 377
  pixels have been inverted. Hamming distortion is used.\\
&&&\\
\picbox{\exppicwidth}{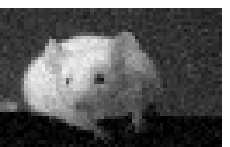}&The
same picture of a mouse, but now zero mean Gaussian noise with
$\sigma=8$ has been added to each pixel. Euclidean distortion is
used; the distortion to the original mouse is $391.1$.%
&\parbox[t]{\exppicwidth}{\tiny Beauty, real beauty, ends2wheresan
intellectual expressoon begins. IntellHct isg in itself a mMde
ofSexggeration, an$\backslash$ destroys theLharmony of n
face. [\ldots]\\\\\\\hbox to\exppicwidth{\hfill\footnotesize\textit{(See Figure~\ref{fig:dorian_gray})}\hfill}}&A
corrupted quotation from Chapter~1 of \emph{The Picture of Dorian
  Gray}, by Oscar Wilde. The 733 byte long fragment was created by
performing 68 random insertions, deletions and replacements of
characters in the original text. Edit distortion is used. The rest of
chapters one and two of the novel are given to the program as side
information.\\ 
\end{tabular}
\caption{The four objects that are subjected to rate-distortion analysis.}
\label{fig:objects}
\end{figure*}

We have subjected four objects to our program. The following
considerations have influenced our choice of objects:
\begin{itemize}
\item Objects should not be too complex, allowing our program to find
  a good approximation of the distortion-rate curve. We found that
  the running time of the program seems to depend mostly on the
  complexity of the original object; a compressed size of 20,000 bits
  seemed to be about the maximum our program could handle within a
  reasonable amount of time, requiring a running time of the order of
  weeks on a desk-top computer.
\item To check that our method really is general, objects should be
  quite different: they should come from different object domains, for
  which different distortion functions are appropriate, and they
  should contain structure at different levels of complexity.
\item Objects should contain primary structure and regularities
that are distinguishable and compressible by a block sorting compressor
  such as the one we use. Otherwise, we may no longer hope that
  the compressor implements a significant approximation of the
  Kolmogorov complexity. For instance, we would not expect our program
  to do well on a sequence of digits from the binary expansion of the
  number $\pi$.
\end{itemize}
With this in mind, we have selected the objects listed in
Figure~\ref{fig:objects}.

\medskip\noindent In each experiment, as time progressed the program
in Appendix~\ref{sect.ga}
found less and less improvements per iteration, but the set of
candidate solutions, called the {\em pool} in the Appendix,
 never stabilized completely. Therefore we interrupted each
experiment when (a) after at least one night of computation, the pool
did not improve a lot, and (b) for all intuitively good models
$y\in\representations$ that we could conceive of a priori, the
algorithm had found an $y'$ in the pool with $y'\lte y$ according
to (\ref{eq.leq}). For example, in
each denoising experiment, this test included the original, noiseless
object. In the experiment on the mouse without added noise, we also
included the images that can be obtained by reducing the number of
grey levels in the original with an image manipulation
program. Finally for the greyscale images we included a number of
objects that can be obtained by subjecting the original object to
JPEG2000 compression at various quality levels.

The first experiment illustrates how algorithmic rate-distortion
theory may be applied to lossy compression problems, and it
illustrates how for a given rate, some features of the image are
preserved while others can no longer be retained. We compare the
performance of our method to the performance of JPEG and JPEG2000 at
various quality levels. Standard JPEG images were encoded using the
ImageMagick version 6.2.2; profile information was stripped. JPEG2000
images were encoded to jpc format with three quality levels using
NetPBM version 10.33.0; all other options are default. For more
information about these software packages refer to~\cite{ImageMagick}
and~\cite{NetPBM}.

The other three experiments are concerned with denoising. Any model
that is output by the program can be interpreted as a denoised version
of the input object. We measure the denoising success of a model $y$
as $d(x',y)$, where $x'$ is the original version of the input object
$x$, before noise was added. We also compare the denoising results to
those of other denoising algorithms:

\begin{enumerate}
\item BayesShrink denoising\cite{Chang2000}. BayesShrink is a
  popular wavelet-based denoising method that is considered to work
  well for images.
\item Blurring (convolution with a Gaussian kernel). Blurring works
  like a low-pass filter, eliminating high frequency information such
  as noise. Unfortunately other high frequency features of the image,
  such as sharp contours, are also discarded.
\item Naive denoising. We applied a naive denoising algorithm to the
  noisy cross, in which each pixel was inverted if five or more out
  of the eight neighbouring pixels were of different colour.
\item Denoising based on JPEG2000. Here we subjected the noisy input
  image to JPEG2000 compression at different quality levels. We then
  selected the result for which the distortion to the original image
  was lowest.
\end{enumerate}

\subsection{Names of Objects}
To facilitate description and discussion of the experiments we will
adopt the following naming convention. Objects related to the
experiments with the mouse, the noisy cross, the noisy mouse and the
Wilde fragment, are denoted by the symbols {\mouse}, {\cross},
{\nmouse} and {\wilde} respectively. A number of important objects in
each experiment are identified by a subscript as follows. For
$\obj\in\{\mouse,\cross,\nmouse,\wilde\}$, the input object, for which
the rate-distortion function is approximated by the program, is called
\os{\obj}{in}, which is sometimes abbreviated to \obj. In the
denoising experiments, the input object is always constructed by
adding noise to an original object. The original objects and the noise
are called \os{\obj}{orig} and \os{\obj}{noise} respectively. If
Hamming distortion is used, addition is carried out modulo 2, so that
the input object is in effect a pixelwise exclusive OR of the original
and the noise. In particular, $\os{\cross}{in}$ equals
$\os{\cross}{orig}$ XOR $\os{\cross}{noise}$. The program outputs the
reduction of the gene pool, which is the set of considered models. Two
important models are also given special names: the model within the
gene pool that minimizes the distortion to \os{\obj}{orig} constitutes
the best denoising of the input object and is therefore called
\os{\obj}{best}, and the minimal sufficient statistic as described in
Section~\ref{sec:sufstat} is called \os{\obj}{mss}. Finally, in the
denoising experiments we also give names to the results of the
alternative denoising algorithms. Namely, \os{\cross}{naive} is the
result of the naive denoising algorithm applied to the noisy cross,
\os{\nmouse}{blur} is the convolution of {\nmouse} with a Gaussian
kernel with $\sigma=0.458$, \os{\nmouse}{bs} is the denoising
result of the BayesShrink algorithm, and \os{\nmouse}{jpeg2000} is the
image produced by subjecting {\nmouse} to JPEG2000 compression at the
quality level for which the distortion to \os{\nmouse}{orig} is
minimized.

\section{Results and Discussion}\label{sec:resdisc}
We will occasionally use terminology from Appendix~\ref{sect.ga}, but
such references can safely be glossed over on first reading.  After
running for some time on each input object, our program outputs the
reduction of a pool \pool, which is interpreted as a set of
models. For each experiment, we report a number of different
properties of these sets.  Since we are interested in the
rate-distortion properties of the input object $x=\os{\obj}{in}$, we
plot the approximation of the distortion-rate function of each input
object: $d_x(r)=\min\{d(x,y):y\in\representations,K(y)\le
r\}\approx\min\{d(x,y):y\in \hbox{trd}(\pool),L_A(y)\le r\}$, where
$L_A$ denotes the codelength for an object under our compression
algorithm. Such approximations of the distortion-rate function are
provided for all four experiments.  For the greyscale images we also
plot the distortion-rate approximation that is achieved by JPEG2000
(and in Figure~\ref{fig:mouse} also ordinary JPEG) at different
quality levels. Here, the rate is the codelength achieved by
JPEG(2000), and the distortion is the Euclidean distortion to
\os{\obj}{in}. We also plot the codelength function as discussed in
Section~\ref{sec:sufstat}. Minimal sufficient statistics can be
identified by locating the minimum of this graph.

\subsection{Lossy Compression}\label{sec:lossy_compression}

\subsubsection*{Experiment 1: \mouse ouse (Euclidean distortion)}
\begin{figure*}[!ht]
  \subfigbottomskip=0cm 
  \hbox to \textwidth{%
    \subfigure[$r\!=\!163.0$, $d\!=\!2210.0$]{\smi{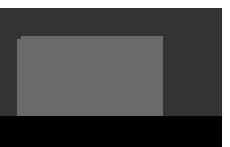}\label{fig:mlca}}\hfill
    \subfigure[$r\!=\!437.8$, $d\!=\!1080.5$]{\smi{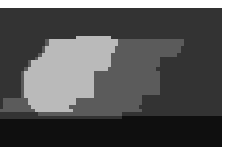}\label{fig:mlcb}}
  }\medskip
    \hbox to \textwidth{%
      \subfigure[$r\!=\!976.6$, $d\!=\!668.9$]{\smi{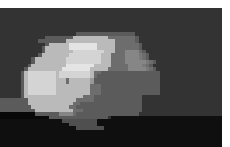}\label{fig:mlcc}}\hfill%
      \subfigure[$r\!=\!1242.9$, $d\!=\!546.4$]{\smi{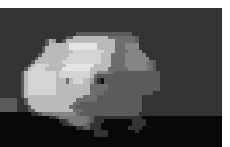}\label{fig:mlcd}}
    }\medskip
  \hbox to \textwidth{%
    \subfigure[$r\!=\!1676.6$, $d\!=\!406.9$]{\smi{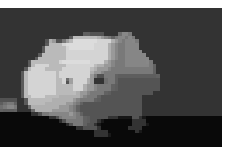}\label{fig:mlce}}\hfill%
    \subfigure[$r\!=\!2324.5$, $d\!=\!298.9$]{\smi{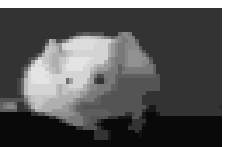}\label{fig:mlcf}}
  }\medskip
  \hbox to \textwidth{%
    \subfigure[$r\!=\!3190.6$, $d\!=\!203.9$]{\smi{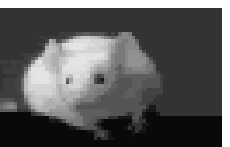}\label{fig:mlcg}}\hfill%
    \subfigure[$r\!=\!7995.0$, $d\!=\!0$]{\smi{mouse_6958.eps}\label{fig:mouse_orig}\label{fig:mlch}}
  }
\caption{Lossy image compression results for {\mouse}
  (Image~\subref{fig:mouse_orig}). Pixel intensities range from 0
  (black) to 255 (white). $r$ denotes the rate (the image's compressed
  length in bits), $d$ denotes the Euclidean distortion to
  \mouse.}
\label{fig:mouse_progression}
\end{figure*}

Our first experiment involved the lossy compression of {\mouse}, a
greyscale image of a mouse. A number of elements of the gene pool are
shown in Figure~\ref{fig:mouse_progression}. The pictures show how at
low rates, the models capture the most important global structures of
the image; at higher rates more subtle properties of the image can be
represented. Figure~\ref{fig:mlca} shows a rough rendering of the
distribution of bright and dark areas in \os{\mouse}{in}. These shapes
are rectangular, which is probably an artifact of the compression
algorithm we used: it is better able to compress images with
rectangular structure than with circular structure. There is no real
reason why a circular structure should be in any way more complex than
a rectangular structure, but most general purpose data compression
software is similarly biased. In \ref{fig:mlcb}, the rate is high
enough that the oval shape of the mouse can be accommodated, and two
areas of different overall brightness are identified. After the number
of grey shades has been increased a little further in~\ref{fig:mlcc},
the first hint of the mouse's eyes becomes visible. The eyes are
improved and the mouse is given paws in~\ref{fig:mlcd}. At higher
rates, the image becomes more and more refined, but the improvements
are subtle and seem of a less qualitative nature.

\begin{figure*}[!ht]
\subfigure[Approximate distortion-rate function]{\centerline{\includegraphics[width=.93\textwidth]{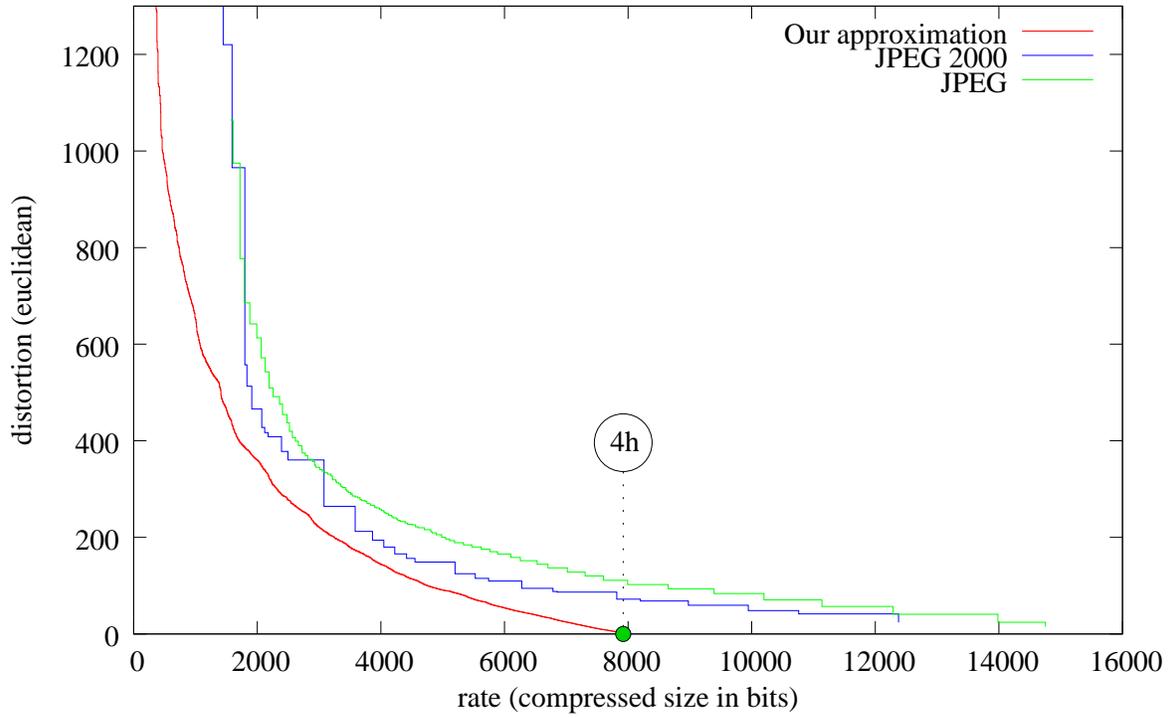}}\label{fig:mouse_dr}}
\subfigure[Approximate codelength function]{\centerline{\includegraphics[width=.93\textwidth]{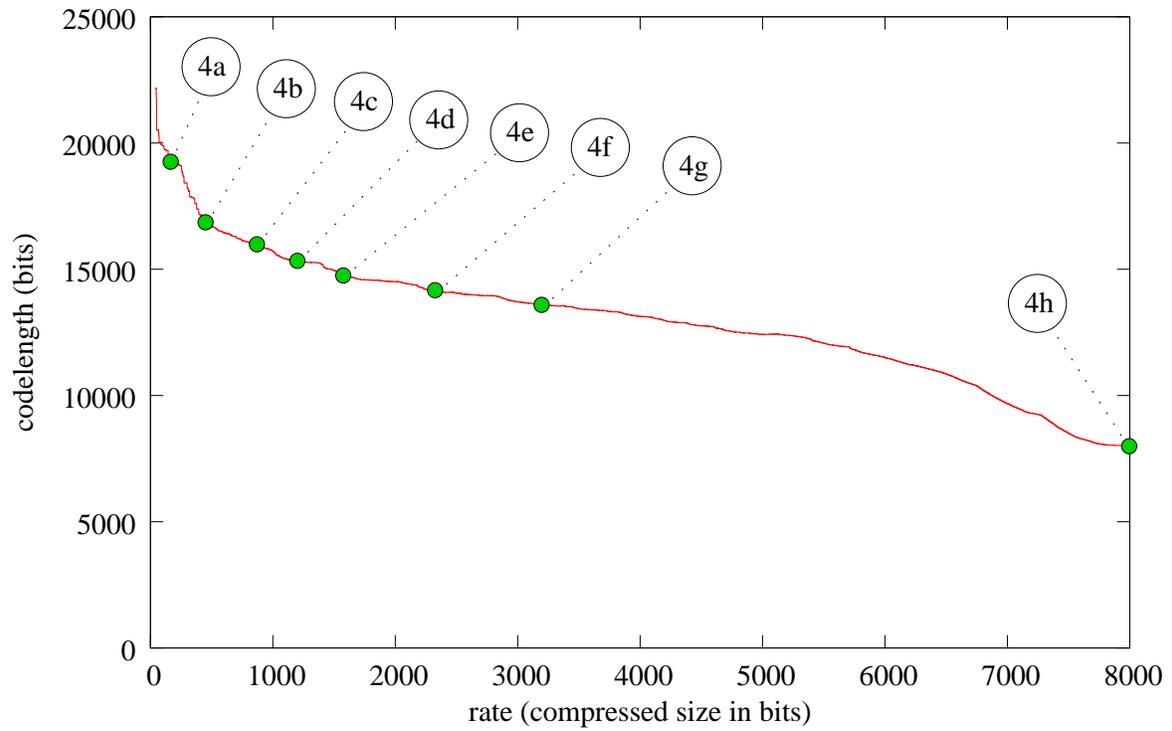}}\label{fig:mouse_cl}}
\caption{Results for the \mouse ouse (Figure~\ref{fig:mouse_progression}).}
\label{fig:mouse}
\end{figure*}

Figure~\ref{fig:mouse_cl} shows that the only sufficient statistic in
the set of models is \os{\mouse}{in} itself, indicating that the image
hardly contains any noise. It also shows the rates that correspond to
the models that are shown in Figure~\ref{fig:mouse_progression}. By
comparing these figures it can be clearly seen that the image quality
only starts to deteriorate significantly after more than half of the
information in \os{\mouse}{in} has been discarded. Note that this is
not a statement about the compression ratio, where the size is related
to the size of the \emph{uncompressed} object. For example,
$\os{\mouse}{in}$ has an uncompressed size of $64\cdot 40\cdot
8=20480$ bits, and the representation in Figure~\ref{fig:mlcg} has a
compressed size of $3190.6$ bits. This representation therefore
constitutes compression by a factor of $20480/3190.6=6.42$, which is
substantial for an image of such small size. At the same time the
amount of \emph{information} is reduced by a factor of
$7995.0/3190.6=2.51$.

\clearpage

\clearpage
\subsection{Denoising}\label{sec:denoising}
For each denoising experiment, we report a number of important
objects, a graph that shows the approximate distortion-rate function
and a graph that shows the approximate codelength function.  In the
distortion-rate graph we plot not only the distortion to \os{\obj}{in}
but also the distortion to \os{\obj}{orig}, to visualise the denoising
success at each rate.

In interpreting these results, it is important to realise that only
the reported minimal sufficient statistic and the results of the
BayesShrink and naive denoising methods can be obtained without
knowledge of the original object -- the other objects \os{\obj}{best},
\os{\obj}{jpeg2000} and \os{\obj}{blur} require selecting between a
number of alternatives in order to optimise the distortion to
\os{\obj}{orig}, which can only be done in a controlled
experiment. Their performance may be better than what can be achieved
in practical situations where \os{\obj}{orig} is not known.

\subsubsection*{Experiment 2: Noisy \cross ross (Hamming distortion)}
\begin{figure*}[!ht]
\subfigbottomskip=0cm 
\subfigure[Approximate distortion-rate function and denoising success
  function. Marked are \os{\cross}{best} (left) and \os{\cross}{in}
  (right). We have $L_A(\os{\cross}{in})=3178.6$,
  $d(\os{\cross}{orig},\os{\cross}{in})=377$, $L_A(\os{\cross}{best})=260.4$ and
  $d(\os{\cross}{orig},\os{\cross}{best})=3$.]{\centerline{\includegraphics[width=.8\textwidth]{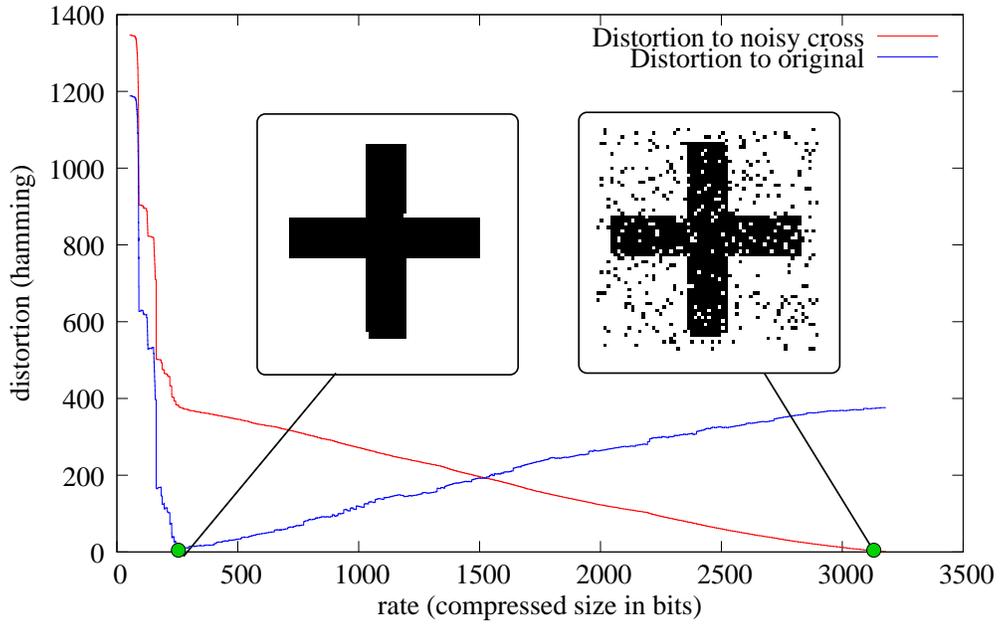}}\label{fig:cross_dr}}

\subfigure[Approximate codelength function. \os{\cross}{mss} is marked. In this case we have
  $\os{\cross}{mss}=\os{\cross}{best}$.]{\centerline{\includegraphics[width=.8\textwidth]{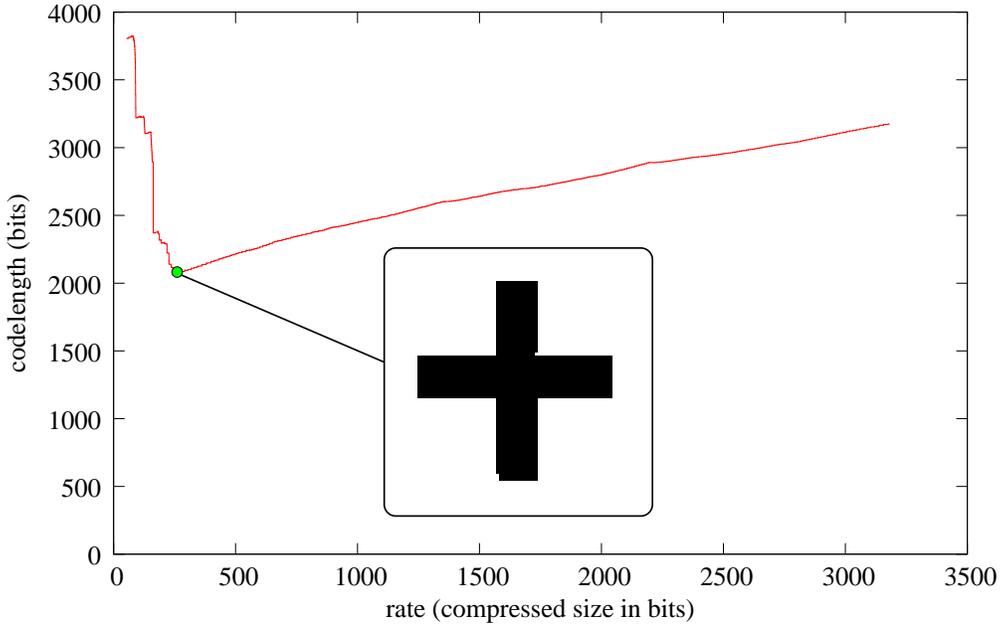}}\label{fig:cross_cl}}

\centerline{%
  \subfigure[$\os{\cross}{naive}$. \hfill$L_A(\os{\cross}{naive})=669.2$,
  $d(\os{\cross}{orig},\os{\cross}{naive})=40$, $d(\os{\cross}{in},\os{\cross}{naive})=389$.]{\qquad\qquad\includegraphics[width=0.24\textwidth]{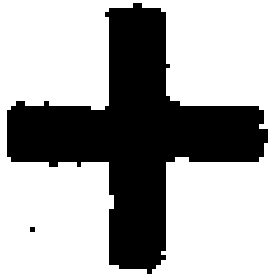}\qquad\qquad\label{fig:cross_naive}}\qquad%
  \subfigure[$\os{\cross}{naive}-\os{\cross}{orig}$]{\qquad\qquad\includegraphics[width=0.24\textwidth]{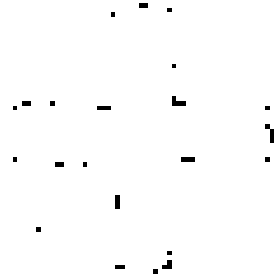}\qquad\qquad\label{fig:cross_naive_residue}}}

\caption{Results for the noisy \cross ross (top right inset).}

\label{fig:cross}
\end{figure*}

In the first denoising experiment we approximated the distortion-rate
function of a monochrome cross \os{\cross}{orig} of very low
complexity, to which artificial noise was added to obtain
\os{\cross}{in}. Figure~\ref{fig:cross} shows the result; the
distortion to the noiseless cross is displayed in the same graph. The
best denoising \os{\cross}{best} has a distortion of only $3$ to the
original \os{\cross}{orig}, which shows that the distortion-rate
function indeed separates structure and noise extremely well in this
example. Figure~\ref{fig:cross_cl} shows the codelength function for
the noisy cross; the minimum on this graph is the minimal sufficient
statistic \os{\cross}{mss}. In this low complexity example, we have
$\os{\cross}{mss}=\os{\cross}{best}$, so the best denoising is not
only very good in this simple example, but it can also be
identified.

We did not subject {\cross} to BayesShrink or blurring because those
methods are unsuitable to monochrome images. Therefore we used the
extremely simple, ``naive'' denoising method that is described in
Section~\ref{sec:experiments} on this specific image instead. The
result is shown in Figure~\ref{fig:cross_naive}; while it does remove
most of the noise, 40 errors remain which is a lot more than those
incurred by the minimal sufficient
statistic. Figure~\ref{fig:cross_naive_residue} shows that all errors
except one are close to the contours of the cross. This illustrates
how the naive algorithm is limited by its property that it takes only
the local neighbourhood of each pixel into account, it cannot
represent larger structures such as straight lines.

\subsubsection*{Experiment 3: \nmouse oisy mouse (Euclidean distortion)}
\begin{figure*}[!ht]
\subfigbottomskip=0cm 
\centerline{%
  \subfigure[$\os{\nmouse}{orig}=\os{\mouse}{in}$;\hfill\break$r\!=\!7995.0$, $d\hbox{\subref{fig:mouse_orig2}}=0$, $d\hbox{\subref{fig:mouse_noise}}=392.1$]{\smj{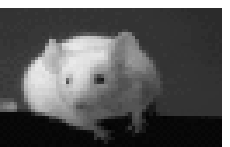}\label{fig:mouse_orig2}}\qquad\qquad
  \subfigure[\os{\nmouse}{in};\hfill\break$r\!=\!16699.7$, $d\hbox{\subref{fig:mouse_orig2}}=392.1$,
  $d\hbox{\subref{fig:mouse_noise}}=0$]{\smj{mouse_noise_1.eps}\label{fig:mouse_noise}}}

\centerline{%
  \subfigure[\os{\nmouse}{best};\hfill\break$r\!=\!3354.4$,
    $d\hbox{\subref{fig:mouse_orig2}}\!=\!272.2$, $d\hbox{\subref{fig:mouse_noise}}\!=\!483.4$]{\smj{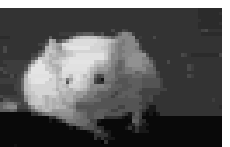}\label{fig:mouse_best}}\qquad\qquad
  \subfigure[$\os{\nmouse}{best}-\os{\nmouse}{orig}$]{\smj{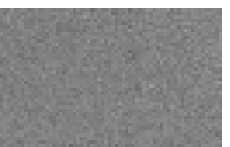}\label{fig:mouse_best_residue}}}

 \centerline{%
 \subfigure[\os{\nmouse}{mss};\hfill\break$r\!=\!1969.8$, $d\hbox{\subref{fig:mouse_orig2}}\!=\!337.0$,
    $d\hbox{\subref{fig:mouse_noise}}\!=\!474.7$]{\smj{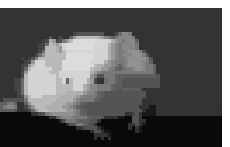}\label{fig:mouse_mdl}}\qquad\qquad
  \subfigure[$\os{\nmouse}{mss}-\os{\nmouse}{orig}$]{\smj{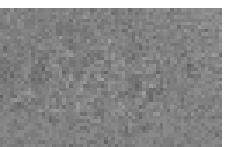}}}

\centerline{%
  \subfigure[\os{\nmouse}{blur}, Gaussian kernel $\sigma=0.458$; $r\!=\!14117$, $d\hbox{\subref{fig:mouse_orig2}}\!=\!291.2$, $d\hbox{\subref{fig:mouse_noise}}\!=\!260.4$]{\smj{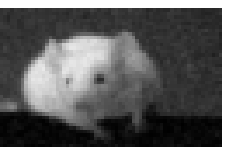}\label{fig:mouse_blur}}\qquad\qquad
  \subfigure[$\os{\nmouse}{blur}-\os{\nmouse}{orig}$]{\smj{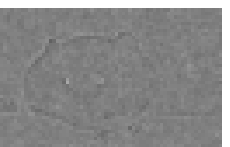}\label{fig:mouse_blur_residue}}}

\centerline{%
  \subfigure[\os{\nmouse}{jpeg2000};\hfill\break $r\!=\!2704$, $d\hbox{\subref{fig:mouse_orig2}}\!=\!379.8$, 
    $d\hbox{\subref{fig:mouse_noise}}\!=\!444.4$]{\smj{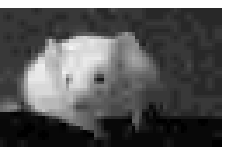}\label{fig:mouse_jpeg}}\qquad\qquad
  \subfigure[$\os{\nmouse}{jpeg2000}-\os{\nmouse}{orig}$]{\smj{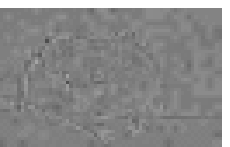}}}

\caption{Denoising results for the \nmouse oisy Mouse. Pixel
intensities range from 0 (black) to 255 (white). $r$ denotes the rate
(the image's compressed length in bits), $d(\hbox{label})$ denotes the
Euclidean distortion to image (label).}
\label{fig:mice}
\end{figure*}

\begin{figure*}[!ht]
\subfigure[Approximate distortion-rate graph for the \nmouse oisy
  Mouse. \os{\nmouse}{best}, \os{\nmouse}{jpeg2000} and
  \os{\nmouse}{in} are marked.]{\centerline{\includegraphics[width=.93\textwidth]{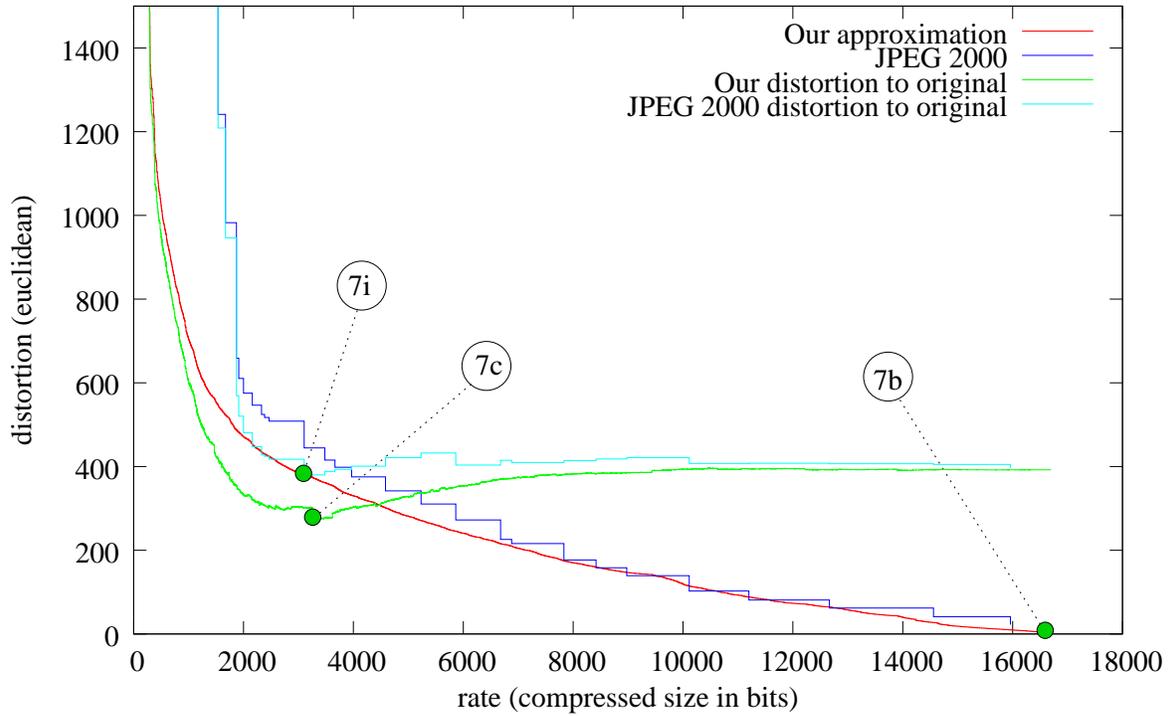}}}
\subfigure[Approximate codelength function. The minimal sufficient
  statistic \os{\nmouse}{mss} is marked.]{\centerline{\includegraphics[width=.93\textwidth]{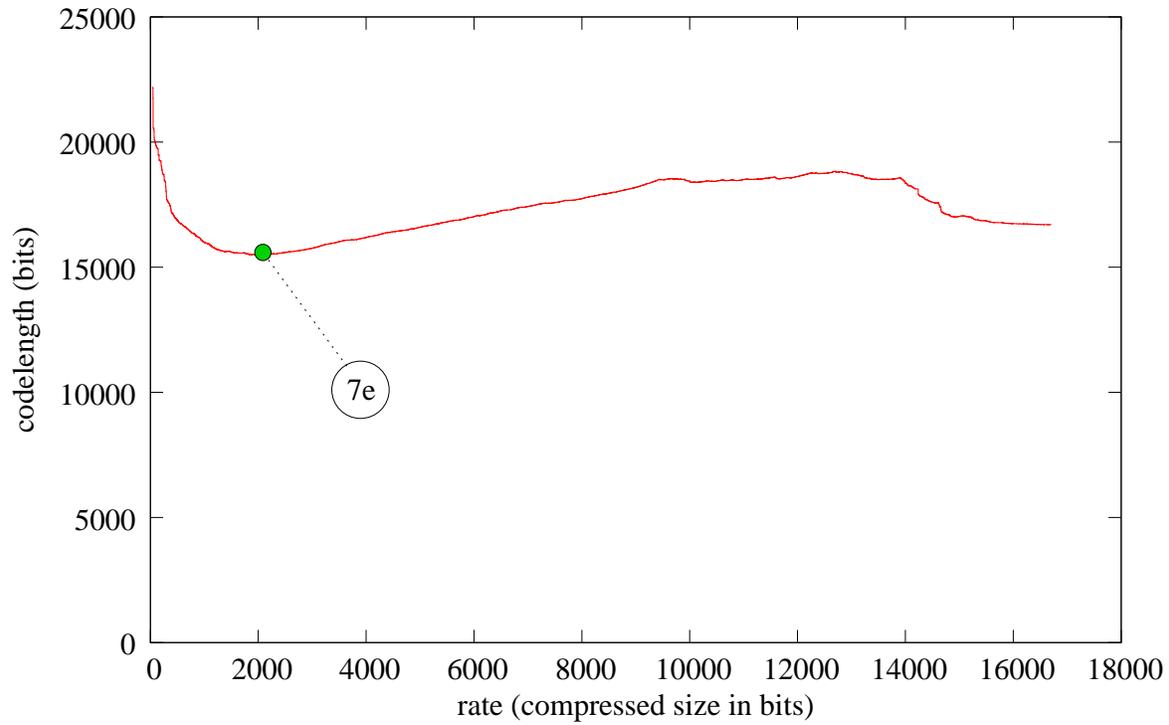}\label{fig:mousenoise_cl}}}
\caption{Results for the \nmouse oisy Mouse}
\label{fig:mousenoise}
\end{figure*}

The noisy mouse poses a significantly harder denoising problem, where
the total complexity of the input \os{\nmouse}{in} is more than five
times as high as for the noisy cross. The graphs that show the
approximation to the distortion-rate function and the codelength
function are in Figure~\ref{fig:mousenoise}, below we discuss the
approximations that are indicated on the graphs with references to
Figure~\ref{fig:mice}.

Figure~\ref{fig:mouse_noise} shows the input object \os{\nmouse}{in};
it was constructed by adding noise to the original noiseless image
\os{\nmouse}{orig}=\os{\mouse}{in}. We have displayed the denoising
results which were obtained in three different ways. In
Figure~\ref{fig:mouse_best} we have shown \os{\nmouse}{best}, the best
denoised object from the gene pool. Visually it appears to resemble
\os{\nmouse}{orig} quite well, but it might be the case that there is
structure in \os{\nmouse}{orig} that is lost in the denoising
process. Because human perception is perhaps the most sensitive
detector of structure in image data, we have shown the difference
between \os{\nmouse}{best} and \os{\nmouse}{orig} in
Figure~\ref{fig:mouse_best_residue}. We would expect any significant
structure in the original image that is lost in the denoising process,
as well as structure that is not present in the original image, but is
somehow introduced as an artifact of the denoising procedure, to
become visible in this residual image. In the case of
\os{\nmouse}{best} we cannot make out any particular features in the
residual.

We have done the same for the minimal sufficient statistic
(Figure~\ref{fig:mouse_mdl}). The result also appears to be a quite
successful denoising, although it is clearly of lower complexity than
the best one. This is also visible in the residual, which still does
not appear to contain much structure, but darker and lighter patches
are definitely discernible. Apparently the difference between
\os{\nmouse}{in} and \os{\nmouse}{mss} does contain some structure,
but is nevertheless coded more efficiently using a literal description
than using the given compression algorithm. We think that the fact
that the minimal sufficient statistic is of lower complexity than the
best possible denoising result should therefore again be attributed to
inefficiencies of the compressor.

For comparison, we have also denoised {\nmouse} using the alternative
denoising method BayesShrink and the methods based on blurring and
JPEG2000 as described in Section~\ref{sec:experiments}. We found that
BayesShrink does not work well for images of such small size: the
distortion between \os{\nmouse}{bs} and \os{\nmouse}{in} is only 72.9,
which means that the input image is hardly effected at all. Also,
\os{\nmouse}{bs} has a distortion of 383.8 to \os{\nmouse}{orig},
which is hardly less than the distortion of 392.1 achieved by
\os{\nmouse}{in} itself.

\bigskip%
{\footnotesize (Continued after Figures~\ref{fig:cross}--\ref{fig:mousenoise}.)}

\clearpage

Blurring-based denoising yields much better results:
\os{\nmouse}{blur} is the result after optimisation of the size of the
Gaussian kernel. Its distortion to \os{\nmouse}{orig} lies in-between
the distortions achieved by \os{\nmouse}{mss} and \os{\nmouse}{best},
but it is different from those objects in two important
respects. Firstly, \os{\nmouse}{blur} remains much closer to
\os{\nmouse}{in}, at a distortion of 260.4 instead of more than 470,
and secondly, \os{\nmouse}{blur} is much less compressible by
$L_A$. (To obtain the reported size of 14117 bits we had to switch on
the averaging filter, as described in Section~\ref{sec:compressor}.)
These observations are at present not well
understood. Figure~\ref{fig:mouse_blur_residue} shows that the
contours of the mouse are somewhat distorted in \os{\nmouse}{blur};
this can be explained by the fact that contours contain high frequency
information which is discarded by the blurring operation as we
remarked in Section~\ref{sec:experiments}.

The last denoising method we compared our results to is the one based
on the JPEG2000 algorithm. Its performance is clearly inferior to our
method visually as well as in terms of rate and distortion. The result
seems to have undergone a smoothing process similar to blurring which
introduces similar artifacts in the background noise, as is clearly
visible in the residual image. As before, the comparison may be
somewhat unfair because JPEG2000 was not designed for the purpose of
denoising, might optimise a different distortion measure and is much
faster.

\subsubsection*{Experiment 4: Oscar \wilde ilde fragment (edit distortion)}
The fourth experiment, in which we analyze \os{\wilde}{in}, a
corrupted quotation from Oscar Wilde, shows that our method is a
general approach to denoising that does not require many domain
specific assumptions. \os{\wilde}{orig}, \os{\wilde}{in} and
\os{\wilde}{mss} are depicted in Figure~\ref{fig:dorian_gray}, the
distortion-rate approximation, the distortion to \os{\wilde}{orig} and
the three part codelength function are shown in
Figure~\ref{fig:wilde}.  We have trained the compression algorithm by
supplying it with the rest of Chapters~1 and 2 of the same novel as
side information, to make it more efficient at compressing fragments
of English text. We make the following observations regarding the
minimal sufficient statistic:

\begin{itemize}
\item In this experiment, $\os{\wilde}{mss}=\os{\wilde}{best}$ so the
minimal sufficient statistic separates structure from noise extremely
well here.
\item The distortion is reduced from 68 errors to only 46 errors. 26
  errors are corrected ($\blacktriangle$), 4 are introduced
  ($\blacktriangledown$), 20 are unchanged ($\bullet$) and 22 are
  changed incorrectly ($\bigstar$).
\item The errors that are newly introduced ($\blacktriangledown$) and
  the incorrect changes ($\bigstar$) typically simplify the fragment a
  lot, so that the compressed size may be expected to drop
  significantly. Not surprisingly therefore, many of the symbols
  marked $\blacktriangledown$ or $\bigstar$ are deletions, or
  modifications that create a word which is different from the
  original, but still correct English. The following table lists
  examples of the last category:
  \begin{center}
    {\footnotesize
    \begin{tabular}{clll}
      Line&\os{\wilde}{orig}&\os{\wilde}{in}&\os{\wilde}{mss}\\
    \hline
    3&\texttt{or}&\texttt{Nor}&\texttt{of}\\
    3&\texttt{the}&\texttt{Ghe}&\texttt{he}\\
    4&\texttt{any}&\texttt{anL}&\texttt{an}\\
    4&\texttt{learned}&\texttt{JeaFned}&\texttt{yearned}\\
    4&\texttt{course}&\texttt{corze}&\texttt{core}\\
    5&\texttt{then}&\texttt{ehen}&\texttt{when}\\
    7&\texttt{he}&\texttt{fhe}&\texttt{the}\\
  \end{tabular}}
  \end{center}
  Since it would be hard for \emph{any} general-purpose 
 mechanical method (that 
does not incorporate a specialized full 
linguistic model of English) to determine that 
  these changes are incorrect, we should not be surprised to find a
  number of errors of this kind.
\end{itemize}

\subsubsection*{Side Information}
\enlargethispage{0.5cm}
Figure~\ref{tab:sideinfo} shows that the compression performance is
significantly improved if we provide side information to the
compression algorithm, and the improvement is typically larger if (1)
the amount of side information is larger, or (2) if the compressed
object is more similar to the side information. Thus, by giving side
information, correct English prose is recognised as ``structure''
sooner and a better separation between structure and noise is to be
expected. The table also shows that if the compressed object is in
some way different from the side information, then adding more side
information will at some point become counter-productive, presumably
because the compression algorithm will then use the side information
to build up false expectations about the object to be compressed,
which can be costly.

While denoising performance probably increases if the amount of side
information is increased, it was infeasible to do so in this
implementation. Recall from Section~\ref{sec:practical} that he
conditional Kolmogorov complexity $K(y|z)$ is approximated by
$L_A(zy)-L_A(z)$. The time required to compute this is dominated by
the length of $z$ if the amount of side information is much larger
than the size of the object to be compressed. This could be remedied
by using a compression algorithm $A'$ that operates sequentially from
left to right, because the state of such an algorithm can be cached
after processing the side information $z$; computing $L_{A'}(zy)$
would then be a simple matter of recalling the state that was reached
after processing $z$ and then processing $y$ starting from that
state. Many compression algorithms, among which Lempel-Ziv compressors
and most statistical compressors, have this property; our approach
could thus be made to work with large quantities of side information
by switching to a sequential compressor but we have not done this.

\clearpage

\begin{figure*}[!ht]
\begin{center}
\begin{tabular}{lrrr}
Side information&$\!\!\!\!\!\!L_A(\os{\wilde}{orig})$&$L_A(\os{\wilde}{mss})$&$L_A(\os{\wilde}{in})$\\
\hline
None&3344.1&3333.7&3834.8\\
Chapters 1,2 (57 kB)&1745.7&1901.9&3234.5\\
Whole novel (421 kB)&1513.6&1876.5&3365.9\\
\end{tabular}
\end{center}
\caption{Compressed size of models for different amounts of side
  information. \os{\wilde}{orig} is never included in the side
  information. We do not let \os{\wilde}{mss} vary with side
  information but keep it fixed at the object reported in
  Figure~\ref{fig:wilde_mdl}.}\label{tab:sideinfo}
\end{figure*}

\begin{figure*}[!ht]

\subfigcapmargin=2em
\subfigcapskip=0.4em

\subfigure[\os{\wilde}{orig}, the original text]{%
\vbox{\hyphenpenalty=10000
\quad Beauty, real beauty, ends where an intellectual expression
begins. Intellect is in itself a\break mode of exaggeration, and destroys
the harmony of any face.  The moment one sits down to think, one
becomes all nose, or all forehead, or something horrid.  Look at the
successful\break men in any of the learned professions. How perfectly
hideous they are!   Except, of course, in the Church.  But then in the
Church they don't think.  A bishop keeps on saying at the age\break of
eighty what he was told to say when he was a boy of eighteen, and as a
natural consequence he always looks absolutely delightful.  Your
mysterious young friend, whose name you have never told me, but whose
picture really fascinates  me, never thinks.  I feel quite sure of
that.}\label{fig:wilde_orig}}

\bigskip

\subfigure[\os{\wilde}{in}, the corrupted version of the fragment. At
  68 randomly selected positions characters have been inserted,
  deleted or modified. New and replacement characters are drawn
  uniformly from ASCII symbols 32--126.]{%
  \vbox{\hyphenpenalty=10000%
  \quad Beauty, real beauty, ends2wheresan intellectual expressoon
  begins. IntellHct isg in itself  a mMde ofSexggeration,
  an$\backslash$ destroys theLharmony of n face. :The m1ment one
  sits down\break to ahink@ one becomes jll noe\^{ } Nor all
  forehbead, or something hNrrid. Look a Ghe successf$\backslash$l men
  in anL of te JeaFned professions. How per\}ectly tideous 4they
  re6  Except, of corze, in7 the Ch4rch. BuP ehen in the Church
  they dol't bthink. =A bishop keeps on saying at the age of eighty
  what he was told to say wh"n he was aJb4y of  eighten, and sja
  natural cnsequence\break fhe a(ways looks ab8olstely de[ightfu). Your
  mysterious youngL friend, wPose name you h$\backslash$vo never tld
  me, mut whose picture really fa?scinates Lme,Pnever thinCs.  I feel
  quite surS of that9 }\label{fig:wilde_in}}

\bigskip

\subfigure[$\os{\wilde}{best}=\os{\wilde}{mss}$; it has edit distortion
  46 to the original fragment. Marks indicate the error type:
  {\scriptsize$\blacktriangle$}=correction;
  {\scriptsize$\blacktriangledown$}=new error; $\bullet$=old error;
  {\tiny$\bigstar$}=changed but still wrong. Deletions are represented
  as \markdeletion.]{%
\vbox{\baselineskip=1.9em
    \quad Beauty, real beauty, ends\markchange{-}where\markplus{ }an
  intellectual express\marksame{o}on begins. Intell\markplus{e}ct
  is\markplus{\markdeletion{}} in itself a\break m\markplus{o}de
  of\markplus{ }ex\marksame{\markdeletion{}}ggeration, an\markplus{d}
  destroys the\markplus{ }harmony of
  \marksame{\markdeletion{}}n\marksame{\markdeletion{}}
  face.~\markchange{\markdeletion{}}The m\markplus{o}ment one sits
  down\break to \markplus{t}hink\markchange{\markdeletion{}} one becomes
  \markchange{\markdeletion{}}ll
  no\marksame{\markdeletion{}}e\markchange{\markdeletion{}}
  \markplus{\markdeletion{}}o\markminus{f} all
  fore\markchange{b}e\markminus{\markdeletion{}}d, or something
  h\markchange{i}rrid. Look a\marksame{\markdeletion{}}
  \markchange{\markdeletion{}}he successf\markchange{\markdeletion{}}l\break
  men in an\markchange{\markdeletion{}} of
  t\marksame{\markdeletion{}}e \markchange{y}ea\markplus{r}ned
  pro\markminus{v}essions. How per\markplus{f}ectly \marksame{t}ideous
  \markplus{\markdeletion{}}they
  \marksame{\markdeletion{}}re\marksame{6} Except, of
  co\marksame{\markdeletion{}}r\markchange{\markdeletion{}}e,
  in\break \markplus{\markdeletion{}} the Ch\markchange{a}rch. Bu\markplus{t}
  \markchange{w}hen in the Church they do\marksame{l}'t
  \markplus{\markdeletion{}}think.~\markchange{\markdeletion{}}A\markminus{\markdeletion{}}bishop
  keeps on saying at the age\break of eighty what he was told to say
  wh\markchange{\markdeletion{}}n he was
  a\markchange{\markdeletion{}}b\markchange{s}y
  of eight\marksame{\markdeletion{}}en, and
  \marksame{\markdeletion{}}s\marksame{j}a natural
  c\marksame{\markdeletion{}}nsequence \markchange{t}he
  a\markchange{\markdeletion{}}ways looks
  ab\markplus{s}ol\markplus{u}tely
  de\markchange{\markdeletion{}}ightfu\markplus{l}. Your
  mysterious young\markplus{\markdeletion{}} friend, w\markplus{h}ose
  name you h\markplus{a}v\markplus{e} never
  t\marksame{\markdeletion{}}ld me, \marksame{m}ut whose picture
  really fa\markplus{\markdeletion{}}scinates
  \markplus{\markdeletion{}}me,\markplus{ }never thin\marksame{C}s.  I
  feel quite sur\markchange{\markdeletion{}} of
  that\marksame{9} }\label{fig:wilde_mdl}}

\caption{A fragment of \emph{The Picture of Dorian Gray}, by Oscar
  \wilde ilde.}
\label{fig:dorian_gray}
\end{figure*}

\begin{figure*}[!ht]
\subfigure[Approximate distortion-rate function, \os{\wilde}{best} and
\os{\wilde}{in} are marked.]{\centerline{\includegraphics[width=.93\textwidth]{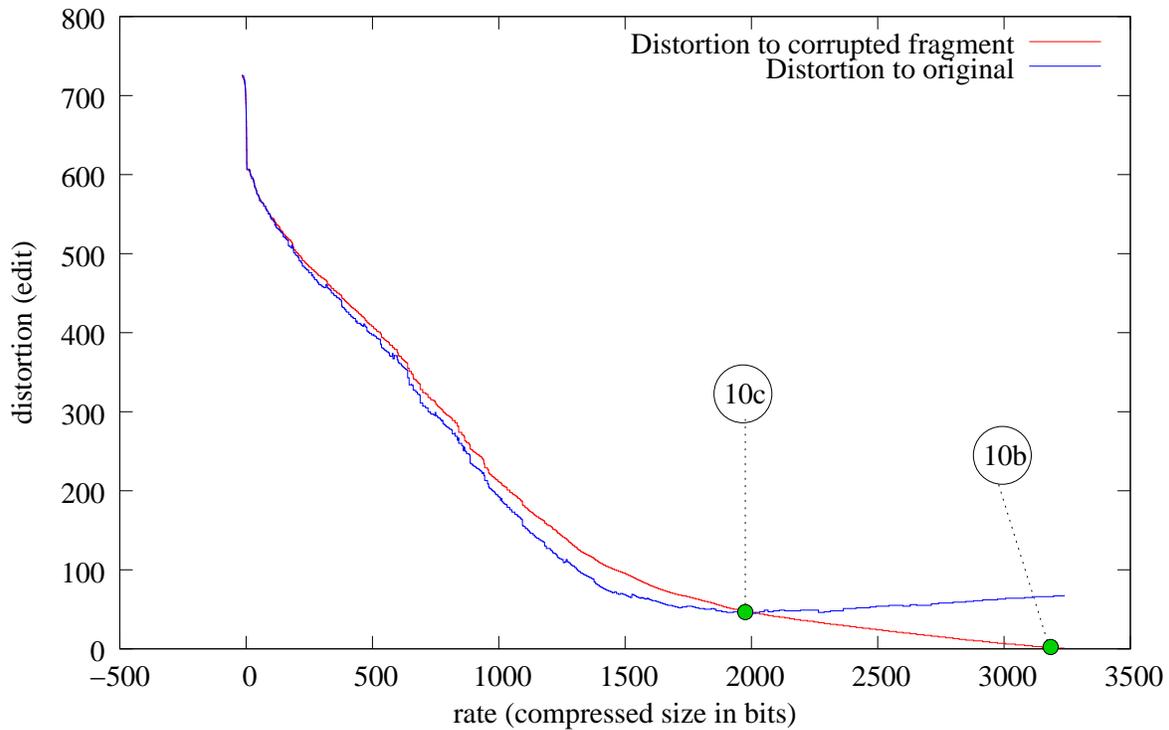}}}
\subfigure[Approximate codelength function, \os{\wilde}{mss} is marked]{\centerline{\includegraphics[width=.93\textwidth]{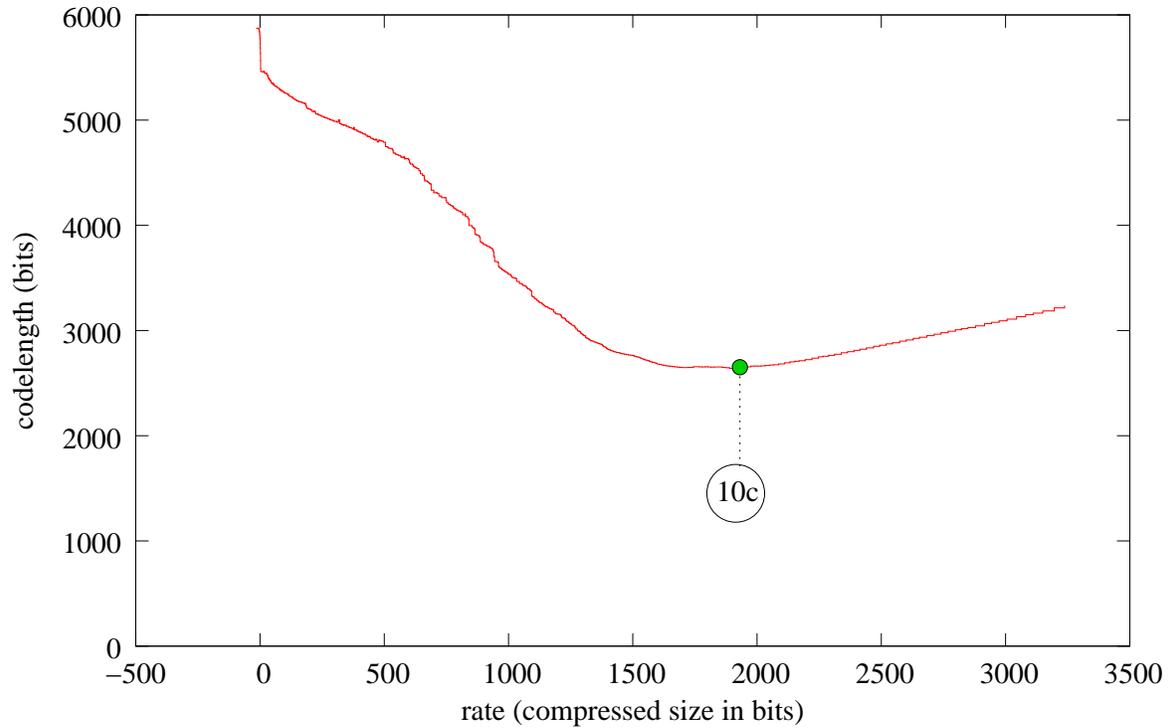}}}
\caption{Results for a fragment of \textit{The Picture of Dorian Gray}
  by Oscar \wilde ilde (also see Figure~\ref{fig:dorian_gray}).\label{fig:wilde}}
\end{figure*}

\clearpage

\section{Quality of the Approximation}\label{sec:quality}
It is easy to see from its definition that the distortion-rate
function must be a non-increasing function of the rate. The
implementation guarantees that our approximation is also
non-increasing. In~\cite{VereshchaginVitanyi2005} it is assumed that
for every $x\in\sourcewords$ there exists a representation
$y\in\representations$ such that $d(x,y)=0$; in the context of this
paper this is certainly true because we have
$\sourcewords=\representations$ and a distortion function which is a
metric. The gene pool is initialised with \os{\obj}{in}, which
always has zero weakness and must therefore remain in the
pool. Therefore at a rate that is high enough to specify $x$, the
distortion-rate function reaches zero.

The shape of the codelength function for an object $x$ is more
complicated. Let $y$ be the representation for which $d(x,y)=0$. In
theory, the codelength can never become less than the complexity of
$y$, and the minimal sufficient statistic witnesses the codelength
function at the lowest rate at which the codelength is equal to the
complexity of $y$. Practically, we found in all denoising experiments
that the total codelength using the minimal sufficient statistic,
$L'_{\os{\obj}{mss}}(\os{\obj}{in})$, is \emph{lower} than the
codelength $L_A(\os{\obj}{in})$ that is obtained by compressing the
input object directly. This can be observed in
Figures~\ref{fig:cross_cl}, \ref{fig:mousenoise_cl} and
\ref{fig:wilde_mdl}. The effect is most clearly visible
in~\ref{fig:cross_cl}, where the separation between structure and
noise is most pronounced.

Our hypothesis is that this departure from the theoretical shape of
the codelength function must be explained by inefficiency of the
compression algorithm in dealing with noise. This is evidenced by the
fact that it needs $2735.7$ bits to encode \os{\cross}{noise}, while
only $\log_2{4096\choose 377}\approx1810$ bits would suffice if the
noise were specified with a uniform code on the set of indices of all
binary sequences with exactly $377$ ones out of $64\cdot 64$ (see
Appendix~\ref{app:dist_hamming}). Similarly,
$L_A(\os{\nmouse}{noise})=14093$, whereas a literal encoding requires
at most $12829$ bits (using the bound in
Appendix~\ref{app:dist_euclid}). 

Another strange effect occurs in Figure~\ref{fig:mousenoise_cl},
where the codelength function displays a strange ``bump'': as the rate
is increased beyond the level required to specify the minimal
sufficient statistic, the codelength goes up as before, but here at
very high rates the codelength starts dropping again.

It is theoretically possible that the codelength function should
exhibit such behaviour to a limited extent. It can be seen
in~\cite{VereshchaginVitanyi2005} that a temporary increase in the
codelength function can occur up to a number of bits that depends on
the so-called \emph{covering coefficient}. Loosely speaking this is
the density of small distortion balls that is required in order to
completely cover a larger distortion ball. The covering coefficient in
turn depends on the used distortion function and the number of
dimensions. It is quite hard to analyze in the case of Euclidean
distortion, so we cannot at present say if theory admits such a large
increase in the codelength function. However, we believe that the
explanation is more mundane in this case. Since the noisy mouse is the
most complex object of the four we experimented on, we fear that this
bump may simply indicate that we interrupted our search procedure too
soon. Quite possibly, after a few years of running the codelength
function would have run straight in between \os{\nmouse}{mss} and
\os{\nmouse}{in}.

Figure~\ref{fig:mouse_dr} shows that our approximation of the
distortion-rate function is somewhat better than the approximation
provided by either JPEG or JPEG2000, although the difference is not
extremely large for higher rates. The probable reason is twofold: on
the one hand, we do not know for which distortion function JPEG(2000)
is optimised, but it might well be something other than Euclidean
distortion. If this is the case, then our comparison is unfair because
our method might well perform worse on JPEG(2000)'s own distortion
measure. On the other hand, JPEG(2000) is very time-efficient: It took
only a matter of seconds to compute models at various different
quality levels, while it took our own algorithm days or weeks to
compute its distortion-rate approximation. Two conclusions can be
drawn from our result. Namely, if the performance of existing image
compression software had been better than the performance of our own
method in our experiments, this would have been evidence to suggest
that our algorithm does not compute a good approximation to the
rate-distortion function. The fact that this is not the case is thus
reassuring. Vice versa, if we assume that we have computed a good
approximation to the algorithmic rate-distortion function, then our
results give a measure of how close JPEG(2000) comes to the
theoretical optimum; our program can thus be used to provide a basis
for the evaluation of the performance of lossy compressors.

\section{Conclusion}\label{sec:conclusion}
Algorithmic rate-distortion provides a good
framework for analysis of large and structured objects.
It is based on Kolmogorov
complexity, which is not computable. We nevertheless attempted to put
this theory into practice by approximating the Kolmogorov complexity
of an object by its compressed size. We also generalized the theory in
order to enable it to cope with side information, which is interpreted
as being available to both the sender and the receiver in a
transmission over a rate restricted channel. We also describe how
algorithmic rate-distortion theory may be applied to lossy compression
and denoising problems.

Finding the approximate rate-distortion function of an individual
object is a difficult search problem. We describe a genetic algorithm
that is very slow, but has the important advantage that it requires
only few assumptions about the problem at hand. Judging from our
experimental results, our algorithm provides a good approximation, as
long as its input object is of reasonably low complexity and is
compressible by the used data compressor. The shape of the approximate
rate-distortion function, and especially that of the associated three
part codelength function, is reasonably similar to the shape that we
would expect on the basis of theory, but there is a striking
difference as well: at rates higher than the complexity of the minimal
sufficient statistic, the three part codelength tends to increase with
the rate, where theory suggests it should remain constant. We expect
that this effect can be attributed to inefficiencies in the
compressor.

We find that the algorithm performs quite well in lossy compression,
with apparently somewhat better image quality than that achieved by
JPEG2000, although the comparison may not be altogether fair.  When
applied to denoising, the minimal sufficient statistic tends to be a
slight underestimate of the complexity of the best possible denoising
(an example of underfitting). This is presumably again due to
inefficiencies in the used compression algorithm.

\bibliography{paper}

\clearpage

\appendix
\section{Appendix: Search Algorithm}
\label{sect.ga}
 We use an almost completely generic procedure to simultaneously
optimise two separate objective functions for objects that are
represented as byte sequences. To emphasize this we will consider the
abstract objective function $g$ wherever possible, rather than the
more concrete rate and distortion functions.

\subsection{Definitions}\label{sec:defs}
We need a number of definitions to facilitate the discussion of the
algorithm and its results below. A finite subset of $\representations$
is called a \emph{pool}. A pool $\pool$ induces a \emph{tradeoff
profile} $p(\pool):=\{g(y):y'\lte y\hbox{~for some $y'$ in
$\pool$}\}$. The \emph{weakness} $w_\pool(y)$ of an object $y\in\pool$
is defined as the number of elements of the pool that are smaller
according to $\lte$. The \emph{(transitive) reduction}
$\hbox{trd}(\pool)$ of a pool $\pool$ is the subset of all elements
with zero weakness. The elements of the reduction of a pool $\pool$
are called \emph{models}.

\subsection{Genetic Algorithm}\label{sec:genalg}
The search algorithm initialises a pool $\pool_0$, which is then
subjected to a process of selection through survival of the fittest:
the pool is iteratively updated by replacing elements with low fitness
(the fitness function is specified below) by new ones, which are
created through either \emph{mutation} (random modifications of
elements) or \emph{crossover} (``genetic'' recombination of pairs of
other candidates). We write $\pool_i$ to denote the pool after $i$
iterations. When the algorithm terminates after $n$ iterations it
outputs the reduction of $\pool_n$.

In what follows we will describe our choices for the important
components of the algorithm: the mechanics of crossover and mutation,
the fitness function and the selection function which specifies the
probability that a candidate is removed from the pool.  In the
interest of reproducibility we faithfully describe all our important
design choices, even though some of them are somewhat arbitrary.

\subsubsection{Crossover}
Crossover (also called recombination) is effected by the following
algorithm. Given two objects $x$ and $y$ we first split them both in
three parts: $x=x_1x_2x_3$ and $y=y_1y_2y_3$, such that the length of
$x_1$ is chosen uniformly at random between $0$ and the length of $x$
and the length of $x_2$ is chosen from a geometric distribution with
mean $5$; the lengths of the $y_i$ are proportional to the lengths of
the $x_i$. We then construct a new object by concatenating
$x_1y_2x_3$.

\subsubsection{Mutation}
The introduction of a mutation operation is necessary to ensure that
the search space is connected, since the closure of the gene pool
under crossover alone might not cover the entire search space. While
we could have used any generic mutation function that meets this
requirement, for reasons of efficiency we have decided to design a
different mutation function for every objective function that we
implemented. This is helpful because some distortion functions (here,
the edit distortion) can compare objects of different sizes while
others cannot: mutation is the means by which introduction of objects
of different size to the pool can be brought about when desirable, or
avoided when undesirable.

The mutation algorithm we use can make two kinds of change. With
probability $1/4$ we make a small random modification using an
algorithm that depends on the distortion function. Below is a table of
the distortion functions and a short description of the associated
mutation algorithms:

\medskip\noindent{\footnotesize\begin{tabular}{l|l}
Distortion&Mutation algorithm\\
\hline Hamming&Sets a random byte to a uniformly random value\\
Euclidean&Adds an ${\cal N}[0;\sigma=10]$ value to a random byte\\
Edit&A random byte is changed, inserted or deleted\\
\end{tabular}}

\medskip With probability $3/4$ we use the following mutation
algorithm instead. It splits the object $x$ into three parts
$x=x_1x_2x_3$ where the length of $x_1$ is chosen uniformly at random
between $0$ and the length of $x$ and the length of $x_2$ is chosen
from a geometric distribution with mean $5$. The mutation is effected
by training a (simplified version of) a third order PPM
model\cite{ClearyWitten1984} on $x_1$ and then replacing $x_2$ with an
equally long sequence that is sampled from the model. The advantage of
this scheme is that every replacement for $x_2$ gets positive
probability, but replacements which have low codelength and
distortion tend to be much more likely than under a uniform
distribution.

\subsubsection{Fitness Function}
In theory, the set of representations that witness the rate-distortion
function does not change under monotonic transformation of either
objective function $g_1$ or $g_2$. We have tried to maintain this
property throughout the search algorithm including the fitness
function, by never using either objective function directly but only
the ordering relation $\lte$. Under such a regime, a very natural
definition of fitness is minus the weakness of the objects with
respect to pool $\pool$.

It has been an interesting mini-puzzle to come up with an efficient
algorithm to compute the weakness of all objects in the pool
efficiently. Our solution is the following very simple algorithm,
which has an $O(n\log n)$ average case running time. It first sorts
all elements of the pool by their value under $g_1$ and then inserts
them in order into a binary search tree in which the elements are
ordered by their value under $g_2$. As an object is inserted into the
tree, we can efficiently count how many elements with lower values for
$g_2$ the tree already contained. These elements are precisely the
objects that have both lower values on $g_1$ (otherwise they would not
appear in the tree yet) and on $g_2$; as such their number is the
desired weakness.

\subsubsection{Selection Function}
It is not hard to see that we have $p(\pool)=p(\hbox{trd}(\pool))$ and
$p(\pool)\subseteq p(\pool\cup\pool')$. Therefore monotonic
improvement of the pool under modification is ensured as long as
candidates with weakness $0$ are never dropped.

\begin{figure}[!ht]
\centerline{\includegraphics[width=\columnwidth]{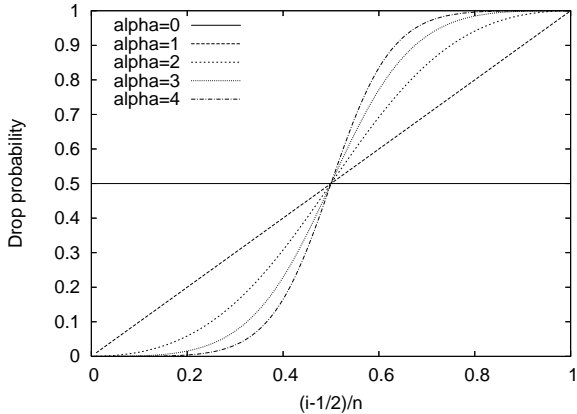}}
\caption{Drop probability as a function of the index}\label{fig:dropprob}
\end{figure}

We drop other candidates with positive probability as follows. Let
$y_1,\ldots,y_n$ be the elements of $\pool$ with nonzero weakness,
ordered such that for $1\le i<j\le n$ we have
$w_\pool(y_i)<w_\pool(y_j)$ or $g_1(y_i)<g_1(y_j)$ if $y_i$ and $y_j$
have the same weakness. We drop candidate $y_i$ from the pool with
probability $1/(1+({n\over i-1/2}-1)^\alpha)$, which is a modified
sigmoid function where $\alpha\in(1,\infty)$ specifies the sharpness
of the transition from probability zero to one. This function is
plotted for different values of $\alpha$ in Figure~\ref{fig:dropprob}.
We used $\alpha=4$ in our experiments.

\section{Appendix: Distortion Spheres}\label{app:distortion}
For each distortion function, we compute or bound from above the size
of a \emph{distortion sphere} around some representation
$y\in\representations$. A sphere with centre $y$ of radius $a$ is a
set of the form $S_y(a):=\{x\in\sourcewords:d(x,y)=a\}$.

Distortion spheres are important in the analysis because they enable
us to link the distortion to the amount of discarded information.

\subsection{The Size of a Hamming Distortion Sphere}\label{app:dist_hamming}
Hamming distortion can be defined when
$\sourcewords=\representations=\Sigma^n$, sequences of $n$ symbols
from a finite alphabet $\Sigma$. Let $y\in\representations$ be an
object of length $n$; there is a bijection between the elements of
$S_y(a)$ and all possible ways to replace $a$ symbols in $y$ with
different symbols from the alphabet. Thus the size of the sphere can
be calculated as ${n\choose a}(|\Sigma|-1)^a$.

\subsection{The Size of a Euclidean Distortion Sphere}\label{app:dist_euclid}
Our variety of Euclidean distortion requires that
$\sourcewords=\representations={\mathbb{Z}}^n$, the set of
$n$-dimensional vectors of integers. The size of a Euclidean
distortion sphere around some $y\in\representations$ of length $n$ is
hard to compute analytically. We use an upper bound that is reasonably
tight and can be computed efficiently. First we define
$d(v):=d(v,{\vec0})=\sqrt{\sum_{i=1}^n v_i^2}$ and $S(n,a)$ as the set
$\{v:|v|=n,d(v)=a\}$. We have $x\in S_y(a)\Leftrightarrow x-y\in
S(n,a)$, so it suffices to bound the size of $S(n,a)$. We define:
$$p(\delta|n,a):=ce^{-\delta^2n/2a^2}~\hbox{where}~c=1/\sum_{\delta=-255}^{\delta=255}e^{-\delta^2n/2a^2}$$
$$P(v):=\prod_{i=1}^n p(v_i|n,d(v));$$
\noindent $p(\cdot|n,d(v))$ can be interpreted as a probability mass function on
the individual entries of $v$ (which in our application always lie
between -255 and 255). Therefore $P(v)$ defines a valid probability
mass function on outcomes $v$ in $S(n,a)$. Thus,

\begin{eqnarray*}
1& > &\!\!\!\!\sum_{v\in S(n,a)}P(v)=\sum_{v\in S(n,a)}c^ne^{-(\sum
  \delta_i^2)n/2a^2}\\
& = &\!\!\!\!\sum_{v\in S(n,a)}c^ne^{-n/2}=c^ne^{-n/2}|S(n,a)|.
\end{eqnarray*}
This yields a bound on the size of $S(n,a)$, which is reasonably tight
unless the distortion $a$ is very low. In that case, we can improve
the bound by observing that $v$ must have at least $z=n-d(v)^2$ zero
entries. Let $v'$ be a vector of length $n-z$ that is obtained by
removing $z$ zero entries from $v$. Every $v$ in $S(n,a)$ can be
constructed by inserting $z$ zeroes into $v'$, so we have
$|S_y(a)|=|S(n,a)|\le{n\choose z}|S(n-z,a)|$. The size of $S(n-z,a)$
can be bounded by using the method described before recursively.

\subsection{The Size of an Edit Distortion Sphere}\label{app:dist_edit}
Edit distortion can be defined for spaces
$\sourcewords=\representations=\Sigma^*$ for a finite alphabet
$\Sigma$. We develop an upper bound on the size of the edit distortion
sphere. We can identify any object in $S_y(a)$ by a program $p$ that
operates on $y$, and which is defined by a list of instructions to
copy, replace or delete the next symbol from $y$, or to insert a new
symbol. We interpret a deletion as a replacement with an empty symbol;
so the replacement operations henceforth include deletions.  Let
$d(p)$ denote the number of insertions and replacements in $p$.
Clearly for all $x\in S_y(a)$, there must be a $p$ such that $p(y)=x$
and $d(p)=d(x,y)=a$. Therefore the size of $S_y(a)$ can be upper
bounded by counting the number of programs with $d(p)=a$. Let $n$ be
the length of $y$. Any program that contains $i$ insertions and that
processes $y$ completely, must be of length $n+i$. The $i$ insertions,
$a-i$ replacements and $n-a+i$ copies can be distributed over the
program in ${n+i\choose i,a-i, n+a-i}$ different ways. For each
insertion and replacement, the number of possibilities is equal to the
alphabet size. Therefore,
\begin{eqnarray*}
|S_y(a)|& \le &|\{p:d(p)=a\}|\\
& \le &|\Sigma|^d\!\!\!\!\!\!\!\!\sum_{i=\max\{0,a-n\}}^a {n+i\choose i, n-a+i,
  a-i}.\end{eqnarray*}
The sphere can be extremely large, so to facilitate
  calculation of the log of the sphere size, as is required in our
  application, it is convenient to relax the bound some more and
  replace every term in the sum by the largest one. Calculation
  reveals that the largest term has
$$i=\floor{{1\over4}\left(2(a-n)+1+\sqrt{4(n^2+n+a^2+a)+1}\right)}.$$

\end{document}